\pdfoutput=1
\documentclass[12pt,a4paper]{article}
\usepackage{ifthen} % for conditional statements
\newboolean{pdflatex}
\setboolean{pdflatex}{true} % False for eps figures 

\newboolean{articletitles}
\setboolean{articletitles}{true} % False removes titles in references

\newboolean{uprightparticles}
\setboolean{uprightparticles}{false} %True for upright particle symbols

\newboolean{inbibliography}
\setboolean{inbibliography}{false} %True once you enter the bibliography

\def\paperauthors{LHCb collaboration} % Leave as is for PAPER and CONF
\def\paperasciititle{Forward Top Pair Production at LHCb} % Set ASCII title here
\def\papertitle{Measurement of forward top pair production in the dilepton channel in $pp$ collisions at $\sqs=13\tev$} % Latex formatted title
\def\paperkeywords{{High Energy Physics}, {LHCb}} % Comma separated list
\def\papercopyright{CERN on behalf of the LHCb collaboration}
\def\paperlicence{CC-BY-4.0}
\def\paperlicenceurl{https://creativecommons.org/licenses/by/4.0/}

\usepackage[top=1in, bottom=1.25in, left=1in, right=1in]{geometry}

\usepackage{microtype}
\usepackage{lineno}  % for line numbering during review
\usepackage{xspace} % To avoid problems with missing or double spaces after
\usepackage{caption} %these three command get the figure and table captions automatically small

\usepackage{graphicx}  % to include figures (can also use other packages)
\usepackage{color}
\usepackage{colortbl}
\graphicspath{{./figs/}} % Make Latex search fig subdir for figures

\usepackage{amsmath} % Adds a large collection of math symbols
\usepackage{amssymb}
\usepackage{amsfonts}
\usepackage{upgreek} % Adds in support for greek letters in roman typeset

\newcommand*\patchAmsMathEnvironmentForLineno[1]{%
\expandafter\let\csname old#1\expandafter\endcsname\csname #1\endcsname
\expandafter\let\csname oldend#1\expandafter\endcsname\csname
end#1\endcsname
 \renewenvironment{#1}%
   {\linenomath\csname old#1\endcsname}%
   {\csname oldend#1\endcsname\endlinenomath}%
}
\newcommand*\patchBothAmsMathEnvironmentsForLineno[1]{%
  \patchAmsMathEnvironmentForLineno{#1}%
  \patchAmsMathEnvironmentForLineno{#1*}%
}
\AtBeginDocument{%
\patchBothAmsMathEnvironmentsForLineno{equation}%
\patchBothAmsMathEnvironmentsForLineno{align}%
\patchBothAmsMathEnvironmentsForLineno{flalign}%
\patchBothAmsMathEnvironmentsForLineno{alignat}%
\patchBothAmsMathEnvironmentsForLineno{gather}%
\patchBothAmsMathEnvironmentsForLineno{multline}%
\patchBothAmsMathEnvironmentsForLineno{eqnarray}%
}

\usepackage{hyperxmp}

\usepackage[pdftex,
            pdfauthor={\paperauthors},
            pdftitle={\paperasciititle},
            pdfkeywords={\paperkeywords},
            pdfcopyright={Copyright (C) \papercopyright},
            pdflicenseurl={\paperlicenceurl}]{hyperref}

\usepackage[all]{hypcap} % Internal hyperlinks to floats.

\usepackage{xspace} 
\usepackage{upgreek}

\def\lhcb {\mbox{LHCb}\xspace}

\def\MagUp {\mbox{\em Mag\kern -0.05em Up}\xspace}

\ifthenelse{\boolean{uprightparticles}}%
{

 \def\Ppsi        {\ensuremath{\uppsi}\xspace}

 \def\PDelta      {\ensuremath{\Delta}\xspace}                 
 \def\PXi      {\ensuremath{\Xi}\xspace}                 
 \def\PLambda      {\ensuremath{\Lambda}\xspace}                 
 \def\PSigma      {\ensuremath{\Sigma}\xspace}                 
 \def\POmega      {\ensuremath{\Omega}\xspace}                 
 \def\PUpsilon      {\ensuremath{\Upsilon}\xspace}                 
 
 \def\PB      {\ensuremath{\mathrm{B}}\xspace}                 
                  
 \def\PD      {\ensuremath{\mathrm{D}}\xspace}

 \def\PJ      {\ensuremath{\mathrm{J}}\xspace}                 
 \def\PK      {\ensuremath{\mathrm{K}}\xspace}

 \def\PZ      {\ensuremath{\mathrm{Z}}\xspace}                 
                  
 \def\Pb      {\ensuremath{\mathrm{b}}\xspace}                 
 \def\Pc      {\ensuremath{\mathrm{c}}\xspace}

 \def\Pi      {\ensuremath{\mathrm{i}}\xspace}

 \def\Pq      {\ensuremath{\mathrm{q}}\xspace}

 \def\Pt      {\ensuremath{\mathrm{t}}\xspace}

}
{

 \def\Ppsi        {\ensuremath{\psi}\xspace}                 
                  
 \mathchardef\PDelta="7101
 \mathchardef\PXi="7104
 \mathchardef\PLambda="7103
 \mathchardef\PSigma="7106
 \mathchardef\POmega="710A
 \mathchardef\PUpsilon="7107
                  
 \def\PB      {\ensuremath{B}\xspace}                 
                  
 \def\PD      {\ensuremath{D}\xspace}

 \def\PJ      {\ensuremath{J}\xspace}                 
 \def\PK      {\ensuremath{K}\xspace}

 \def\PZ      {\ensuremath{Z}\xspace}                 
                  
 \def\Pb      {\ensuremath{b}\xspace}                 
 \def\Pc      {\ensuremath{c}\xspace}

 \def\Pi      {\ensuremath{i}\xspace}

 \def\Pq      {\ensuremath{q}\xspace}

 \def\Pt      {\ensuremath{t}\xspace}

}

\makeatletter
\ifcase \@ptsize \relax% 10pt
  \newcommand{\miniscule}{\@setfontsize\miniscule{4}{5}}% \tiny: 5/6
\or% 11pt
  \newcommand{\miniscule}{\@setfontsize\miniscule{5}{6}}% \tiny: 6/7
\or% 12pt
  \newcommand{\miniscule}{\@setfontsize\miniscule{5}{6}}% \tiny: 6/7
\fi
\makeatother

\DeclareRobustCommand{\optbar}[1]{\shortstack{{\miniscule (\rule[.5ex]{1.25em}{.18mm})}
  \\ [-.7ex] $#1$}}

\def\Z      {{\ensuremath{\PZ}}\xspace}

\def\quark     {{\ensuremath{\Pq}}\xspace}
\def\quarkbar  {{\ensuremath{\overline \quark}}\xspace}
\def\qqbar     {{\ensuremath{\quark\quarkbar}}\xspace}

\def\cquark    {{\ensuremath{\Pc}}\xspace}

\def\bquark    {{\ensuremath{\Pb}}\xspace}

\def\tquark    {{\ensuremath{\Pt}}\xspace}
\def\tquarkbar {{\ensuremath{\overline \tquark}}\xspace}
\def\ttbar     {{\ensuremath{\tquark\tquarkbar}}\xspace}

  \def\Kbar    {{\kern 0.2em\overline{\kern -0.2em \PK}{}}\xspace}

\def\KorKbar    {\kern 0.18em\optbar{\kern -0.18em K}{}\xspace}

  \def\Dbar    {{\kern 0.2em\overline{\kern -0.2em \PD}{}}\xspace}

\def\DorDbar    {\kern 0.18em\optbar{\kern -0.18em D}{}\xspace}

\def\Bbar    {{\ensuremath{\kern 0.18em\overline{\kern -0.18em \PB}{}}}\xspace}

\def\BorBbar    {\kern 0.18em\optbar{\kern -0.18em B}{}\xspace}

\def\jpsi     {{\ensuremath{{\PJ\mskip -3mu/\mskip -2mu\Ppsi\mskip 2mu}}}\xspace}

  \def\Y#1S{\ensuremath{\PUpsilon{(#1S)}}\xspace}% no space before {...}!

\def\Lbar        {{\ensuremath{\kern 0.1em\overline{\kern -0.1em\PLambda}}}\xspace}
\def\LorLbar    {\kern 0.18em\optbar{\kern -0.18em \PLambda}{}\xspace}

\def\to                 {\ensuremath{\rightarrow}\xspace}

\newcommand{\as}{{\ensuremath{\alpha_s}}\xspace}

\def\AT#1     {\ensuremath{A_{\mathrm{T}}^{#1}}\xspace}           % 2

\def\C#1      {\ensuremath{\mathcal{C}_{#1}}\xspace}                       % 9
\def\Cp#1     {\ensuremath{\mathcal{C}_{#1}^{'}}\xspace}                    % 7
\def\Ceff#1   {\ensuremath{\mathcal{C}_{#1}^{\mathrm{(eff)}}}\xspace}        % 9  
\def\Cpeff#1  {\ensuremath{\mathcal{C}_{#1}^{'\mathrm{(eff)}}}\xspace}       % 7
\def\Ope#1    {\ensuremath{\mathcal{O}_{#1}}\xspace}                       % 2
\def\Opep#1   {\ensuremath{\mathcal{O}_{#1}^{'}}\xspace}                    % 7

\newcommand{\tev}{\ifthenelse{\boolean{inbibliography}}{\ensuremath{~T\kern -0.05em eV}}{\ensuremath{\mathrm{\,Te\kern -0.1em V}}}\xspace}
\newcommand{\gev}{\ensuremath{\mathrm{\,Ge\kern -0.1em V}}\xspace}
\newcommand{\mev}{\ensuremath{\mathrm{\,Me\kern -0.1em V}}\xspace}
\newcommand{\kev}{\ensuremath{\mathrm{\,ke\kern -0.1em V}}\xspace}
\newcommand{\ev}{\ensuremath{\mathrm{\,e\kern -0.1em V}}\xspace}
\newcommand{\gevc}{\ensuremath{{\mathrm{\,Ge\kern -0.1em V\!/}c}}\xspace}
\newcommand{\mevc}{\ensuremath{{\mathrm{\,Me\kern -0.1em V\!/}c}}\xspace}
\newcommand{\gevcc}{\ensuremath{{\mathrm{\,Ge\kern -0.1em V\!/}c^2}}\xspace}
\newcommand{\gevgevcccc}{\ensuremath{{\mathrm{\,Ge\kern -0.1em V^2\!/}c^4}}\xspace}
\newcommand{\mevcc}{\ensuremath{{\mathrm{\,Me\kern -0.1em V\!/}c^2}}\xspace}

\def\mm   {\ensuremath{\mathrm{ \,mm}}\xspace}

\def\mum  {\ensuremath{{\,\upmu\mathrm{m}}}\xspace}

\def\mub{\ensuremath{{\mathrm{ \,\upmu b}}}\xspace}

\def\invfb   {\ensuremath{\mbox{\,fb}^{-1}}\xspace}

\def\gsim{{~\raise.15em\hbox{$>$}\kern-.85em
          \lower.35em\hbox{$\sim$}~}\xspace}
\def\lsim{{~\raise.15em\hbox{$<$}\kern-.85em
          \lower.35em\hbox{$\sim$}~}\xspace}

\def\sqs   {\ensuremath{\protect\sqrt{s}}\xspace}

\def\ptot       {\mbox{$p$}\xspace}
\def\pt         {\mbox{$p_{\mathrm{ T}}$}\xspace}

\def\geant      {\mbox{\textsc{Geant4}}\xspace}

\def\powheg     {\mbox{\textsc{Powheg}}\xspace}
\def\tell1  {TELL1\xspace}
\def\ukl1   {UKL1\xspace}

\def\akt{anti-\ensuremath{k_{\rm T}}\xspace}
\def\bjet{\ensuremath{b}-jet\xspace}
\def\bjets{\ensuremath{b}-jets\xspace}
\def\mueb{\ensuremath{\mu eb}\xspace}
\def\amcatnlo{\mbox{a\textsc{MC@NLO}}\xspace}
\def\fastjet{\mbox{\textsc{Fastjet}}\xspace}
\def\powheg{\mbox{\textsc{POWHEG}}\xspace}
\def\mcfm{\mbox{\textsc{MCFM}}\xspace}
\def\madspin{\mbox{\textsc{MADSPIN}}\xspace}
\def\pythiapp{\mbox{\textsc{Pythia8}}\xspace}

\def\mub{\ensuremath{\mu b}\xspace}

\usepackage{cite} % Allows for ranges in citations
\usepackage{mciteplus}

\usepackage{longtable} % only for template; not usually to be used in PAPERs

\begin{document}

%%%%%%%%%%%%%%%%%%%%%%%%%
%%%%% Title     %%%%%%%%%
%%%%%%%%%%%%%%%%%%%%%%%%%
\renewcommand{\thefootnote}{\fnsymbol{footnote}}
\setcounter{footnote}{1}

% %%%%%%% CHOOSE TITLE PAGE--------
%\onecolumn
%\input{title-LHCb-INT}
%\input{title-LHCb-ANA}
%\input{title-LHCb-CONF}
% $Id: title-LHCb-PAPER.tex 123283 2018-09-03 08:34:59Z sfarry $
% ===============================================================================
% Purpose: LHCb-PAPER journal paper title page template
% Author: 
% Created on: 2010-09-25
% ===============================================================================

%%%%%%%%%%%%%%%%%%%%%%%%%
%%%%%  TITLE PAGE  %%%%%%
%%%%%%%%%%%%%%%%%%%%%%%%%
\begin{titlepage}
\pagenumbering{roman}

% Header ---------------------------------------------------
\vspace*{-1.5cm}
\centerline{\large EUROPEAN ORGANIZATION FOR NUCLEAR RESEARCH (CERN)}
\vspace*{1.5cm}
\noindent
\begin{tabular*}{\linewidth}{lc@{\extracolsep{\fill}}r@{\extracolsep{0pt}}}
\ifthenelse{\boolean{pdflatex}}% Logo format choice
{\vspace*{-1.5cm}\mbox{\!\!\!\includegraphics[width=.14\textwidth]{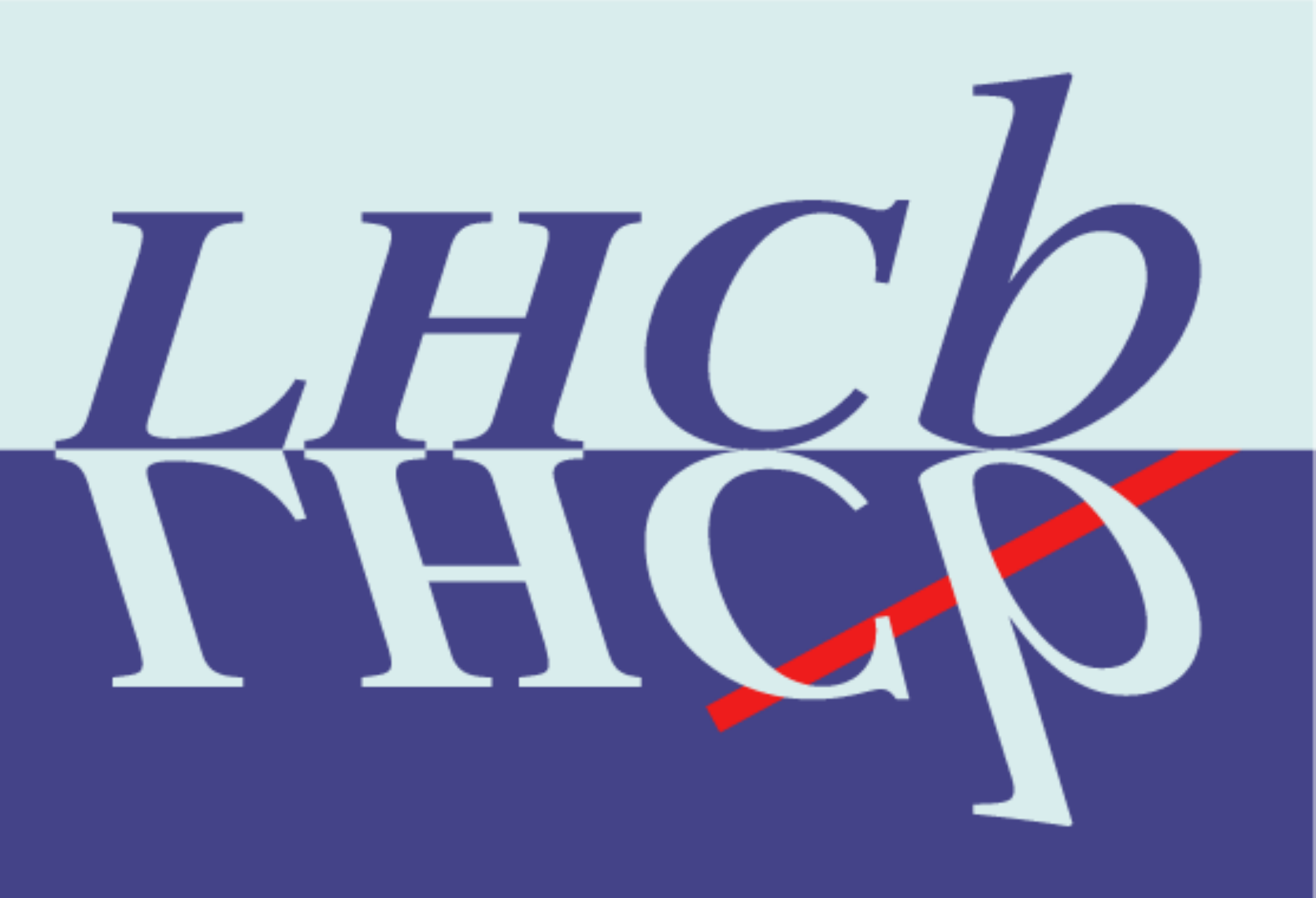}} & &}%
{\vspace*{-1.2cm}\mbox{\!\!\!\includegraphics[width=.12\textwidth]{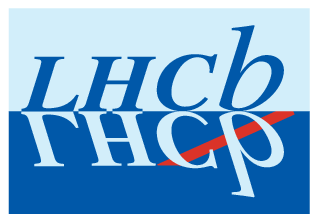}} & &}%
\\
 & & CERN-EP-2018-022 \\  % ID 
 & & LHCb-PAPER-2017-050 \\  % ID 
 & & 3 September 2018 \\ % Date - Can also hardwire e.g.: 23 March 2010
 & & \\
% not in paper \hline
\end{tabular*}

\vspace*{2.0cm}

% Title --------------------------------------------------
{\normalfont\bfseries\boldmath\huge
\begin{center}
% DO NOT EDIT HERE. Instead edit macro in main.tex to keep metadata correct
  \papertitle 
\end{center}
}

\vspace*{1.0cm}

% Authors -------------------------------------------------
\begin{center}
%In the footnote, replace 'paper' by 'Letter' in case of submission to PRL or PLB 
% Edit macro in main.tex to keep metadata correct
\paperauthors\footnote{Authors are listed at the end of this paper.}
\end{center}

\vspace{\fill}

% Abstract -----------------------------------------------
\begin{abstract}
  \noindent
  Forward top quark pair production is studied in $pp$ collisions in the $\mu eb$ final state using a data sample corresponding to an integrated luminosity of 1.93\invfb collected with the LHCb experiment at a centre-of-mass energy of 13~TeV. The cross-section is measured in a fiducial region where both leptons have a transverse momentum greater than 20\gev and a pseudorapidity between 2.0 and 4.5. The quadrature sum of the azimuthal separation and the difference in pseudorapidities, denoted $\Delta R$, between the two leptons must be larger than 0.1. The \bjet axis is required to be separated from both leptons by a $\Delta R$ of 0.5, and to have a transverse momentum in excess of 20\gev and a pseudorapidity between 2.2 and 4.2. The cross-section is measured to be
$$\sigma_{t\bar{t}}= 126\pm19\,(\mathrm{stat})\pm16\,(\mathrm{syst})\pm5\,(\mathrm{lumi})\,\,\mathrm{ fb}$$
where the first uncertainty is statistical, the second is systematic, and the third is due to the luminosity determination. The measurement is compatible with the Standard Model prediction.
\end{abstract}

\vspace*{2.0cm}

\begin{center}
  Published in JHEP 08 (2018) 174
\end{center}

\vspace{\fill}

{\footnotesize 
% Edit macro in main.tex to keep metadata correct
\centerline{\copyright~\papercopyright, licence \href{\paperlicenceurl}{\paperlicence}.}}
\vspace*{2mm}

\end{titlepage}

%%%%%%%%%%%%%%%%%%%%%%%%%%%%%%%%
%%%%%  EOD OF TITLE PAGE  %%%%%%
%%%%%%%%%%%%%%%%%%%%%%%%%%%%%%%%

%  empty page follows the title page ----
\newpage
\setcounter{page}{2}
\mbox{~}
%\newpage
%
%% Author List ----------------------------
%%  You need to get a new author list!
%\input{LHCb_authorlist.tex}
%
%The author list for journal publications is provided by the Membership Committee shortly after 'approval to go to paper' has been given.
%%It will be made available on the page
%%\verb!http://www.physik.uzh.ch/~strauman/forMemCo/LHCb-PAPER-XXXX-XXX/! .
%It will be sent to you by email shortly after a paper number has beens assigned.
%The author list should be included already at first circulation, 
%to allow new members of the collaboration to verify whether they have been included correctly.
%Occasionally a misspelled name is corrected or associated institutions become full members.
%In that case, a new author list will be sent to you.
%In case line numbering doesn't work well after including the authorlist, try moving the \verb!\bigskip! after the last author to a separate line.
%
%
%The authorship for Conference Reports should be ``The LHCb
%  collaboration'', with a footnote giving the name(s) of the contact
%  author(s), but without the full list of collaboration names.

\cleardoublepage

%\twocolumn
% %%%%%%%%%%%%% ---------

\renewcommand{\thefootnote}{\arabic{footnote}}
\setcounter{footnote}{0}

%%%%%%%%%%%%%%%%%%%%%%%%%%%%%%%%
%%%%%  Table of Content   %%%%%%
%%%%%%%%%%%%%%%%%%%%%%%%%%%%%%%%
%%%% Uncomment next 2 lines if desired
%\tableofcontents
%\cleardoublepage

%%%%%%%%%%%%%%%%%%%%%%%%%
%%%%% Main text %%%%%%%%%
%%%%%%%%%%%%%%%%%%%%%%%%%

\pagestyle{plain} % restore page numbers for the main text
\setcounter{page}{1}
\pagenumbering{arabic}

%% Uncomment during review phase. 
%% Comment before a final submission.
%\linenumbers

% You can include short sections directly in the main tex file.
% However, for larger papers it is desirable to split the text into
% several semiautonomous files, which can be revised independently.
% This is especially useful when developing a document in
% collaboration with several people, since then different parts can be
% edited independently.  This type of file organization is shown here.
% 

% !TEX root = main.tex                                                                                                                                                   
% $Id: introduction.tex 121908 2018-07-09 15:06:25Z sfarry $

\section{Introduction}
\label{sec:Introduction}
The production of top quarks at hadron colliders represents an important test of the Standard Model of particle physics (SM). The top quark is the heaviest known fundamental particle and its production and decay properties are sensitive to a number of scenarios beyond the Standard Model. The unique forward acceptance of the LHCb detector allows measurements in a phase space inaccessible to general purpose detectors such as ATLAS and CMS. Top-quark production in this region receives a higher contribution from quark-antiquark (\qqbar) annihilation than in the central region and probes higher values of Bjorken-$x$, where large uncertainties are present in the proton parton distribution functions (PDFs). Precise measurements of top quark production at LHCb can thus be used to constrain PDFs in this region~\cite{Gauld:2013aja}. The greater contribution from quark-initiated production also results in a larger expected charge asymmetry~\cite{Kagan:2011yx,Gauld:2014pxa} in the forward region than in the central region.

The first observation of top-quark production in the forward region was performed by the LHCb collaboration using data corresponding to an integrated luminosity of 3\invfb collected between 2010 and 2012 at centre-of-mass energies of 7 and 8\tev (Run 1)~\cite{LHCb-PAPER-2015-022}. The measurement was performed in the $\mub$ final state, consisting of a muon and a \bjet, where \bjet refers to a jet originating from the fragmentation of a \bquark quark. A precision of 20\% was achieved on the top quark production cross-section. Measurements in this final state have the greatest statistical precision but also suffer from a large background due to the production of a $W$ boson in association with a \bjet. Additionally, this final state does not differentiate between single top quark and top quark pair production. A measurement of top quark production was also performed by the LHCb collaboration in the $\mu b\bar{b}$ and $eb\bar{b}$ final states using the same dataset~\cite{LHCb-PAPER-2016-038}, where the top quark pair (\ttbar) production cross-section was measured with a precision of 40\%. The measurements are in agreement with the SM prediction, calculated to a precision of 25\% to 30\% with \mcfm~\cite{Campbell:2000bg} using the CT10 PDF set~\cite{Lai:2010vv}, and are limited by uncertainties on the $b$-tagging efficiency and the background composition.

In 2015 and 2016, a data sample corresponding to an integrated luminosity of 1.93\invfb was collected by the LHCb experiment at a centre-of-mass energy of 13\tev. The increased energy leads to an increase of a factor of ten in the \ttbar production cross-section within the LHCb acceptance with respect to collisions at a energy of 8\tev~\cite{Gauld:1557385}. The larger cross-section allows access to a number of final states inaccessible in Run 1, including the dilepton final state, where both top quarks decay to a $W$ boson and a \bquark quark, and the $W$ bosons decay leptonically to a lepton and a neutrino. This paper details the first measurement of $t\bar{t}$ production in the \mueb final state at LHCb, where the dilepton channel is partially reconstructed by requiring that a muon, an electron and a \bjet are present in the proton-proton ($pp$) collision, where the leptons are produced by the $W$ boson decay, either directly or through the decay of a tau lepton. This final state yields a high purity with respect to other final states; the selection of two leptons reduces the background from single $W$-boson production and lepton mis-identification, the choice of different flavour leptons suppresses the contribution from the production of \Z bosons, and the $b$-tagged jet reduces the contribution from light jets.

The LHCb detector is introduced in Sec.~\ref{sec:Detector}, the event selection and purity is discussed in Sec.~\ref{sec:selection}, the cross-section calculation is presented in Sec.~\ref{sec:xsec}, the results are given in Sec.~\ref{sec:results} and the conclusions are drawn in Sec.~\ref{sec:conclusions}.

% !TEX root = main.tex                                                                                                                                                   

\section{Detector and simulation}
\label{sec:Detector}

The \lhcb detector~\cite{Alves:2008zz,LHCb-DP-2014-002} is a single-arm forward
spectrometer covering the \mbox{pseudorapidity} range $2<\eta <5$,
designed for the study of particles containing \bquark or \cquark
quarks. The detector includes a high-precision tracking system
consisting of a silicon-strip vertex detector surrounding the $pp$
interaction region, a large-area silicon-strip detector located
upstream of a dipole magnet with a bending power of about
$4{\mathrm{\,Tm}}$, and three stations of silicon-strip detectors and straw
drift tubes placed downstream of the magnet.
The tracking system provides a measurement of momentum, \ptot, of charged particles with
a relative uncertainty that varies from 0.5\% at low momentum to 1.0\% at 200\gev.\footnote{Natural units with $\hbar=c=1$ are used throughout, so that mass and momentum are measured in units of energy.}
The minimum distance of a track to a primary vertex (PV), the impact parameter (IP), 
is measured with a resolution of $(15+29/\pt)\mum$,
where \pt is the component of the momentum transverse to the beam, in\,\gev.
%Different types of charged hadrons are distinguished using information
%from two ring-imaging Cherenkov detectors. 
Photons, electrons and hadrons are identified by a calorimeter system consisting of
scintillating-pad and preshower detectors, an electromagnetic
calorimeter and a hadronic calorimeter. Muons are identified by a
system composed of alternating layers of iron and multiwire
proportional chambers.
The online event selection is performed by a trigger, 
which consists of a hardware stage, based on information from the calorimeter and muon
systems, followed by a software stage, which applies a full event
reconstruction. In this paper, the presence of a muon is used to select candidate events at both stages of the trigger; the hardware stage selects muons with a transverse momentum $p_{\rm T} > 1.76\gev$ and the subsequent software trigger requires that a muon with $\pt > 12\gev$ is present. 

To estimate the trigger, reconstruction and selection efficiencies, to determine background contributions, and to compare the selected data sample to theoretical expectations, simulated $pp$ collisions are generated using
\pythiapp~\cite{Sjostrand:2007gs,Sjostrand:2006za}. The
interaction of the generated particles with the detector, and its response,
are implemented using the \geant
toolkit~\cite{Allison:2006ve} as described in
Ref.~\cite{LHCb-PROC-2011-006}.

Results are compared to theoretical predictions calculated at 
%$\mathcal{O}(\as^2)$
next-to-leading-order (NLO) in perturbative QCD using the \powheg~\cite{Nason:2004rx,Frixione:2007vw,Alioli:2010xd}, \amcatnlo~\cite{Alwall:2014hca} and \mcfm~\cite{Campbell:1999ah,Campbell:2011bn} generators. The \mcfm prediction is provided at fixed order, while the \powheg and \amcatnlo predictions are interfaced with \pythiapp to provide a parton shower. In the case of \amcatnlo, the decay of the top quarks is performed using \madspin~\cite{Frixione:2007zp,Artoisenet:2012st}.  For background studies, samples of single top quark production in association with a $W$ boson are produced with \powheg using both the diagram removal and diagram subtraction schemes~\cite{Re:2010bp}, and samples of $WW$ and $WZ$ boson production are produced using \amcatnlo. For \powheg and \mcfm, the factorisation and normalisation scales are set to the transverse mass of the final state top quarks, while for \amcatnlo, they are set to half the sum of the transverse mass of the final state particles. In all cases, the dynamics of the colliding protons are described by the NNPDF3.0~\cite{Ball:2014uwa} PDF set, and the mass of the top quark is set to 173\gev.

% !TEX root = main.tex                                                                                                                                                   
% $Id: figures.tex 61168 2014-09-25 23:10:50Z roldeman $
% ===============================================================================
% Purpose: including figures in the standard template
% Author: Tomasz Skwarnicki, Ulrik Egede
% Created on: 2010-09-24
% ===============================================================================

\section{Event selection and purity}
\label{sec:selection}
Events containing a high-\pt muon and electron of opposite charge in addition to a high-\pt jet are selected. Muons are identified as reconstructed tracks that are matched to hits in each of the four muon stations furthest downstream, while electrons are identified as tracks that have left large energy deposits in the preshower detector and electromagnetic calorimeter, in addition to small energy deposits in the hadronic calorimeter. The muons and electrons are required to have \pt in excess of 20\gev, a pseudorapidity in the range $2.0<\eta<4.5$, and to be separated by a distance, $\Delta R$, of greater than 0.1 in ($\eta$,$\phi$) space,  where $\phi$ refers to the azimuthal angle. The reconstructed final state particles used as inputs to the jet building are prepared using a particle flow algorithm and clustered using the \akt algorithm as implemented in \fastjet~\cite{Cacciari:2011ma}, with a distance parameter of 0.5. Requirements are placed on the candidate jet in order to reduce the background formed by particles which are either incorrectly reconstructed or produced in additional $pp$ interactions in the same event. The jet is required to have \pt above 20\gev and a pseudorapidity between 2.2 and 4.2. Separation between the jet and each lepton is ensured by requiring that the $\Delta R$ distance between the jet and the leptons is greater than 0.5. A dedicated tagging algorithm is used to select \bjets. The tagger proceeds by building two-body secondary vertices (SVs) using all tracks not associated to any PV in the event, and merging any vertices sharing a common track. A jet is said to be tagged if the event contains an SV with a flight direction satisfying $\Delta R < 0.5$ with respect to the jet axis, where the flight direction is taken to be the vector joining the SV to the primary vertex. More details are given in Ref.~\cite{LHCb-PAPER-2015-016}. Lepton isolation criteria is applied by rejecting events where the transverse component of the vector sum of the momentum of all reconstructed charged particles within $\Delta R<0.5$ of either lepton is greater than 5\gev. The leptons are also required to be consistent with originating from a common primary vertex, satisfying $\mathrm{IP} < 0.04\mm$. After applying the full selection, a total of 44 events are retained in the data sample.

In addition to the signal, several physics processes contribute to the selected data sample, either through the presence of an identical final state, or through the misidentification of one or more of the final state objects. The following background processes are considered.
\begin{itemize}
\item Lepton misidentification where the muon and electron candidates are produced through the misidentification of one or two hadrons. A number of processes can contribute to this background, including QCD multijet production, $W$ and $Z$ production, and $t\bar{t}$ events where only one lepton is produced in the LHCb acceptance.
\item The production of a $Z$ boson in association with a jet contributes either through the $Z\to\tau^+\tau^-$ final state, where the subsequent tau lepton decays produce a final state with a muon and electron, or through the $Z\to\mu^{+}\mu^{-}$ and $Z\to e^+e^-$ final states, where one of the final state leptons is misidentified. The associated jet can either be a genuine \bjet produced in association with the $Z$ boson, or due to the misidentification of a charm or light jet.
\item The production of a single top quark in association with a $W$ boson, known as $W^{\pm}t$ production, contributes an identical final state.
\item Multiboson processes, such as $W^{+}W^{-}$, $W^{\pm}Z$ and $ZZ$ production, give rise to a high \pt muon and electron in the final state with an associated jet.
\end{itemize}

As the background from lepton misidentification produces events containing leptons with both same- and opposite-sign charges, the number of background events is determined by first applying the full selection to the data but requiring that the leptons should have the same charge. The $b$-tagging requirement is also removed to increase the statistical precision. The number of events selected in this same-sign control region is then extrapolated using a factor obtained in an additional background-enriched control sample, where the identification requirements on the electron candidate are reversed. This gives an expectation of 3.5$\,\pm\,$1.9 events, where the uncertainty is due to the combined statistical uncertainties on the selected number of events and the extrapolation factor, which dominate over any expected systematic effects.

The number of $Z\to\tau^+\tau^-$ events is determined by normalising the number of $Z\to\tau^+\tau^-$ events observed in \pythiapp simulation using the ratio of the number of $Z\to\mu^+\mu^-$ events observed in data and simulation. A total of 0.32 $\pm$ 0.03 events are expected, where the uncertainty is obtained by combining the statistical precision of the determination with the uncertainty on the reconstruction and selection efficiency, determined from comparisons between data and simulation. The contributions from misidentification of a lepton in the dimuon and dielectron decay modes of the $Z$ boson are determined using simulation and found to be negligible.

The $W^\pm t$ background is determined using predictions from \powheg calculated in the diagram removal scheme. The predicted number of events is scaled by the efficiency obtained by reweighting the simulated \ttbar sample to match the kinematics of the $W^{\pm}t$ process, yielding an expectation of 1.8$\,\pm\,$0.5 events. The uncertainty is determined by combining the theoretical uncertainties, determined as described in Sec.~\ref{sec:results}, with the difference in the cross-section as calculated using the diagram removal and diagram subtraction schemes. An additional systematic uncertainty is added to account for the differences in the reconstruction and selection efficiency. The contribution from multiboson processes is determined from simulations to be negligible.

\begin{figure}
\includegraphics[width=0.5\textwidth]{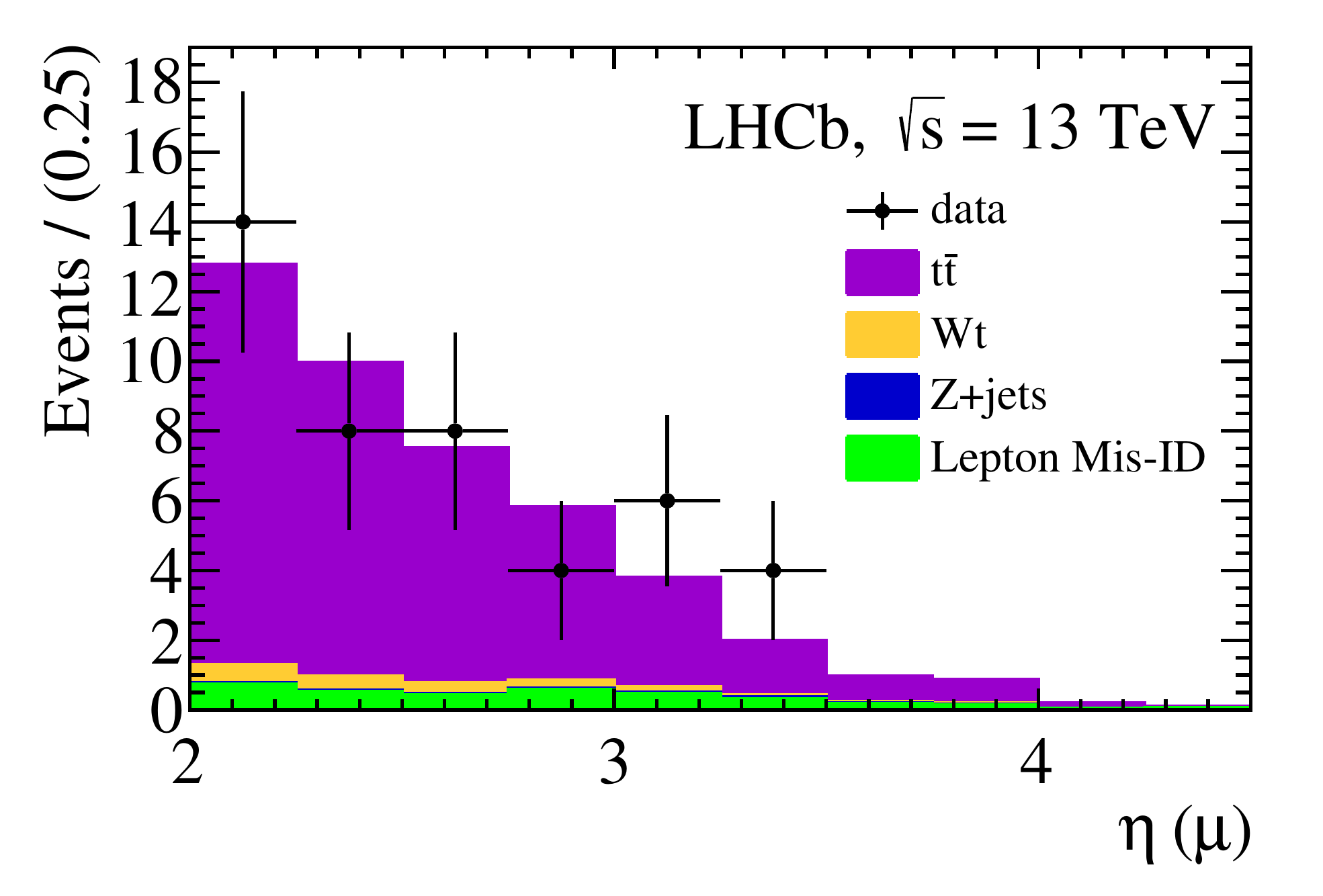}
\includegraphics[width=0.5\textwidth]{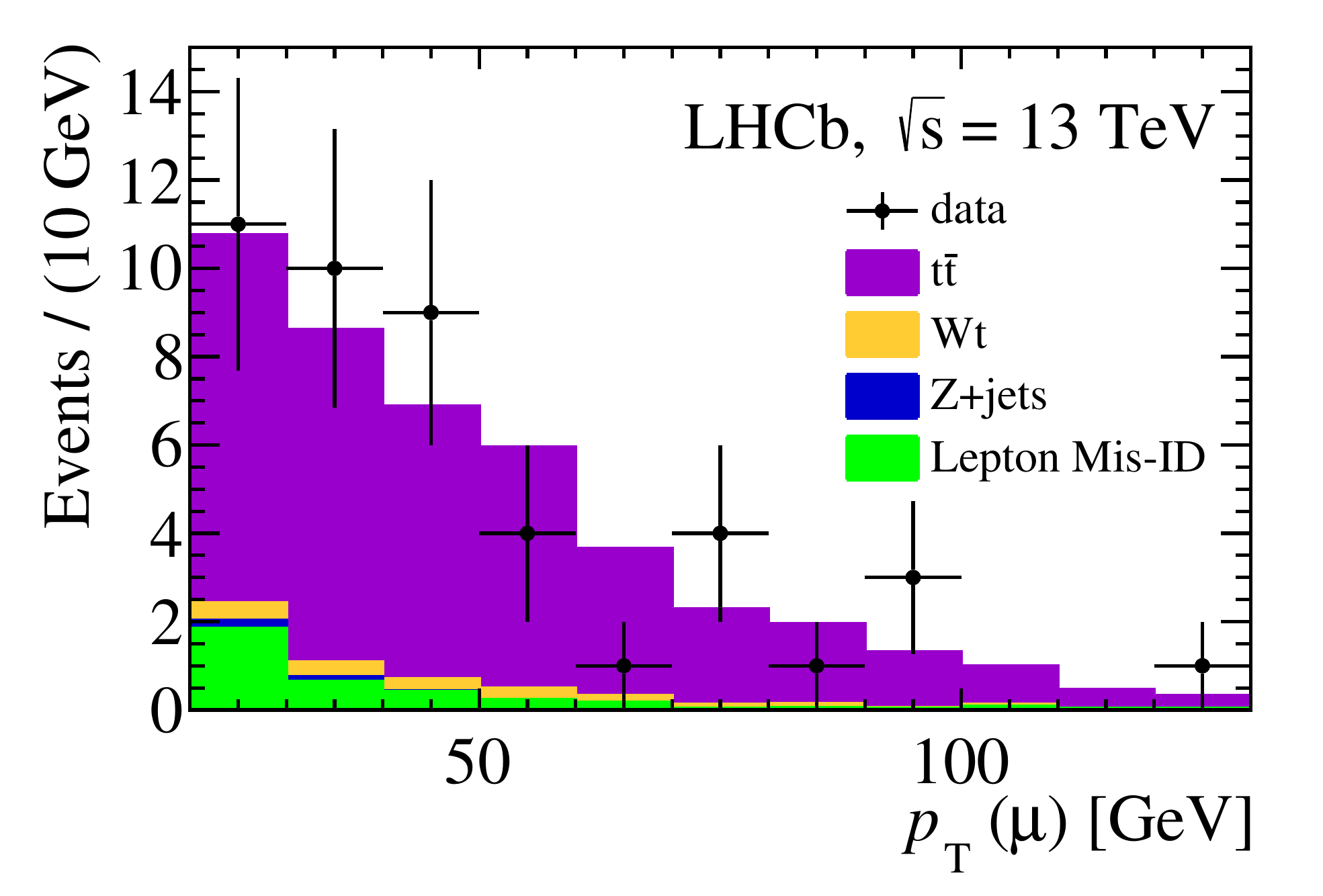}
\caption{The muon (left) $\eta$ and (right) \pt distribution in data compared to the expected contributions. The \ttbar signal yield is determined to be the number of selected events minus the sum of the expected backgrounds. The multiboson background is determined to be negligible.}
\label{fig:muonres}
\end{figure}

\begin{figure}
\includegraphics[width=0.5\textwidth]{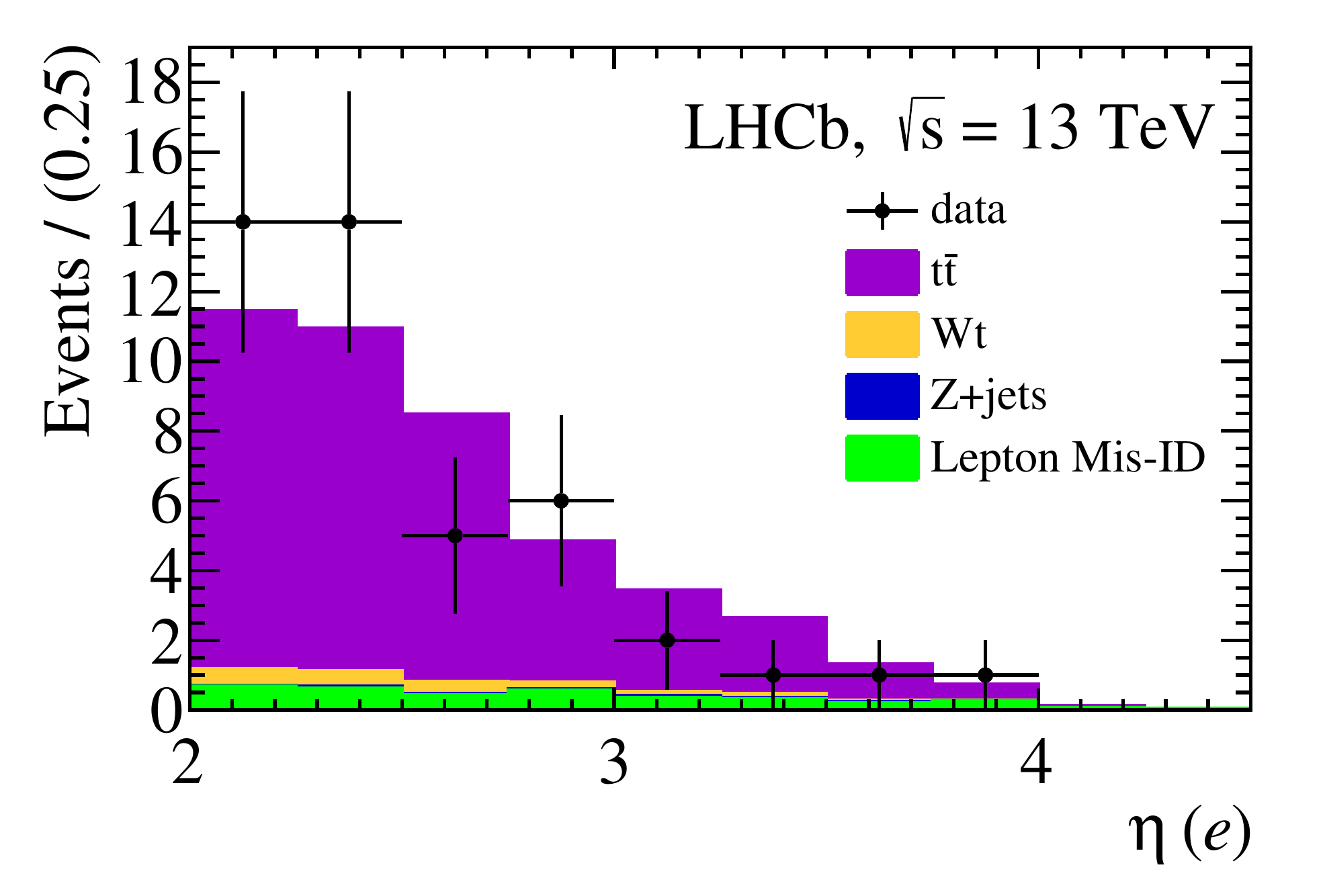}
\includegraphics[width=0.5\textwidth]{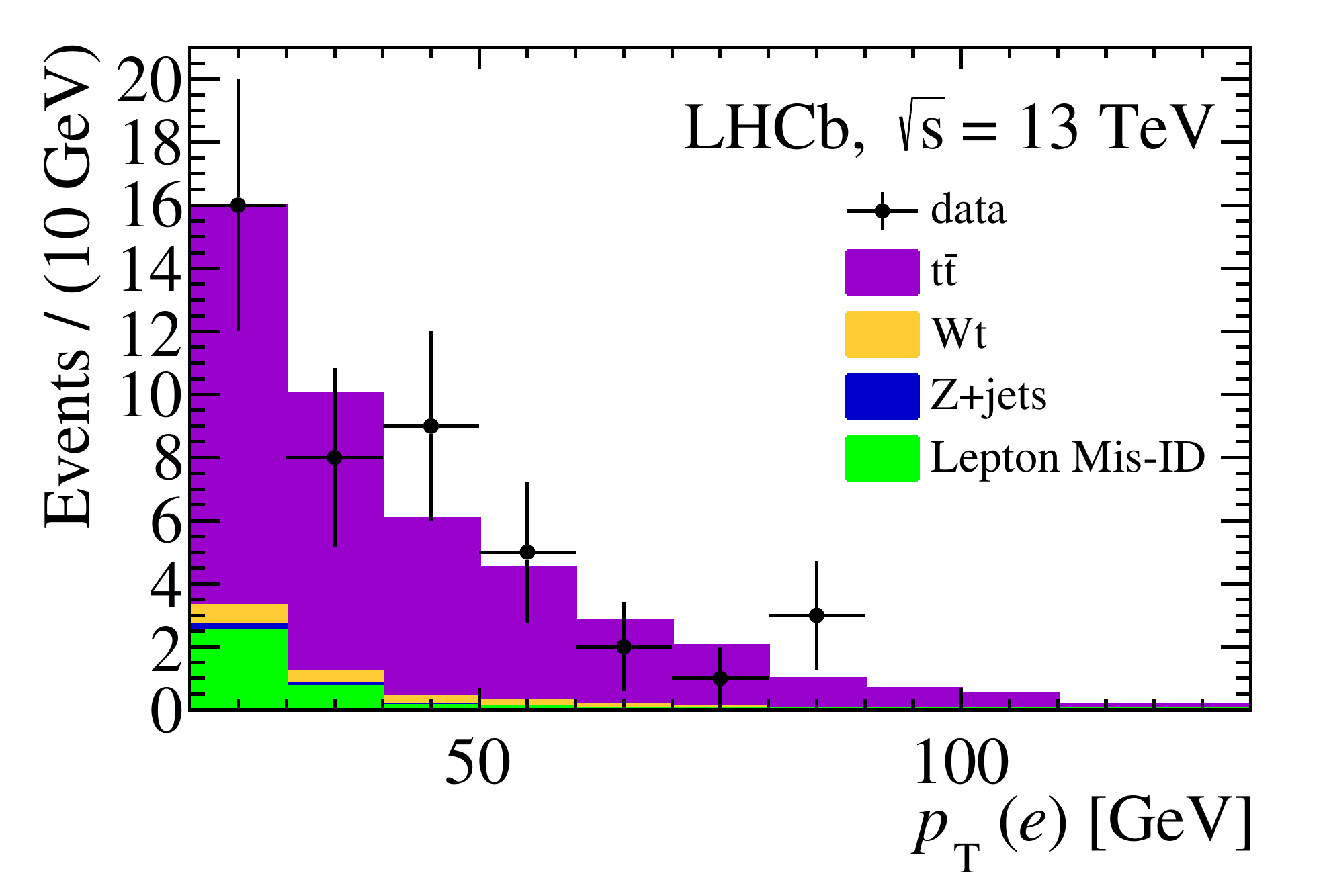}
\caption{The electron (left) $\eta$ and (right) \pt distribution in data compared to the expected contributions. The \ttbar signal yield is determined to be the number of selected events minus the sum of the expected backgrounds. The multiboson background is determined to be negligible.}
\label{fig:electronres}
\end{figure}

\begin{figure}
\includegraphics[width=0.5\textwidth]{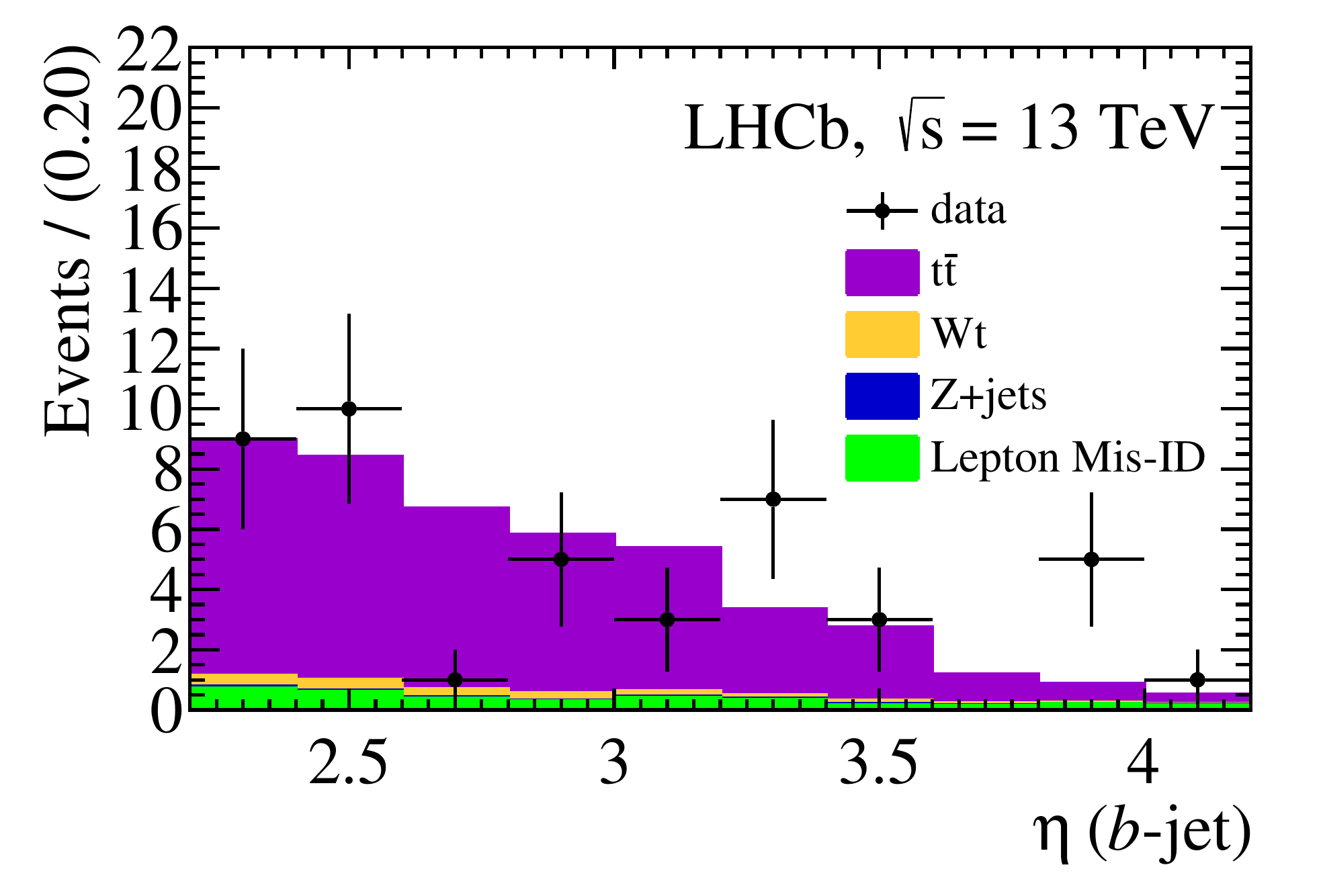}
\includegraphics[width=0.5\textwidth]{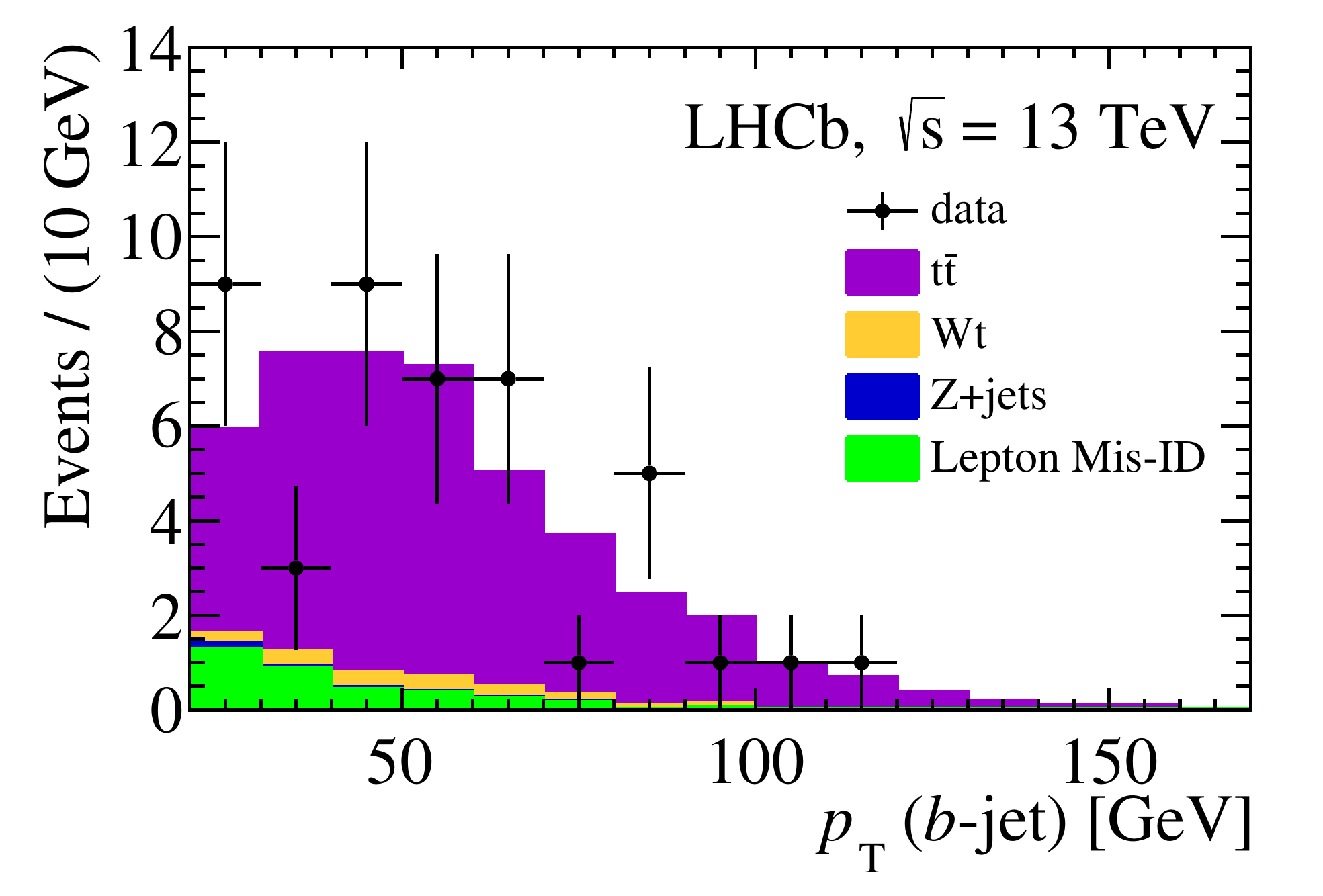}
\caption{The \bjet (left) $\eta$ and (right) \pt distribution in data compared to the expected contributions. The \ttbar signal yield is determined to be the number of selected events minus the sum of the expected backgrounds. The multiboson background is determined to be negligible.}
\label{fig:jetres}
\end{figure}

\begin{figure}
\begin{center}
\includegraphics[width=0.5\textwidth]{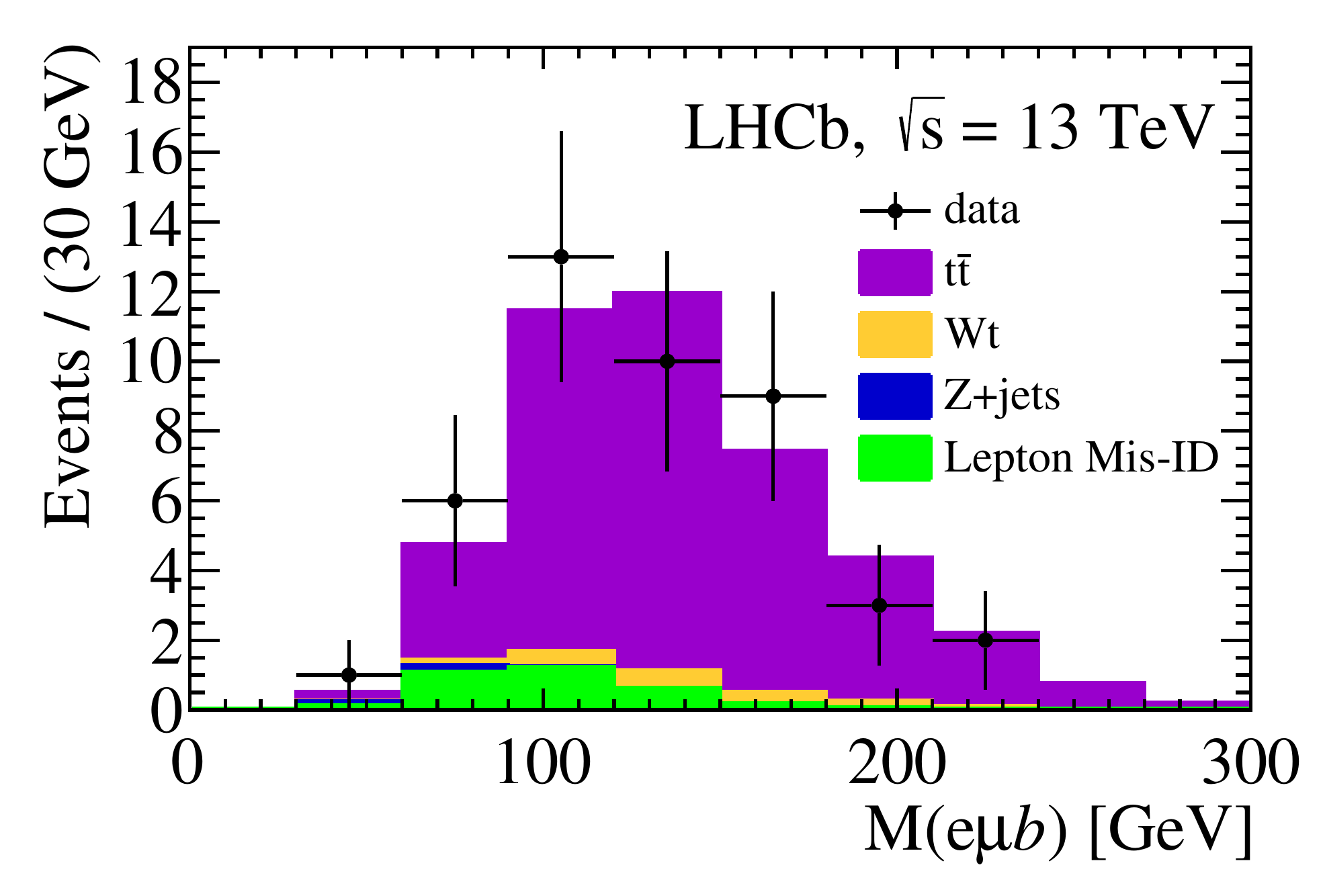}
\end{center}
\caption{The combined invariant mass of the muon, electron and \bjet in data compared to the expected contributions. The \ttbar signal yield is determined to be the number of selected events minus the sum of the expected backgrounds. The multiboson background is determined to be negligible.}
\label{fig:llj_m}
\end{figure}

The total number of expected background events from all sources is 5.6$\,\pm\,$2.0. The total number of selected events is shown as a function of the muon, electron and jet $\eta$ and \pt distributions in Figs.~\ref{fig:muonres},~\ref{fig:electronres} and~\ref{fig:jetres}. The invariant mass of the muon, electron and jet is shown in Fig.~\ref{fig:llj_m}. The expected signal and background contributions are shown, where the signal yield is taken to be the total number of selected events minus the expected background contribution. The lepton misidentification background shape is obtained from the control samples with the electron identification requirements reversed, while all other distributions come from simulation. A reasonable agreement is observed in all distributions.


% !TEX root = main.tex                                                                                                                                                   
\section{Cross-section calculation}
\label{sec:xsec}
The cross-section, $\sigma_\ttbar$, is measured in the fiducial region defined by the \pt, $\eta$, and $\Delta R$ requirements placed on the muon, electron and \bjet candidates, and is calculated using the formula
\begin{equation}
\sigma_\ttbar = \frac{N - N_{\rm bkg}}{\mathcal{L}\cdot\varepsilon}{\cdot\mathcal{F}_{\rm res}},
\end{equation}
where $N$ is the total number of candidates selected in data, $N_{\rm bkg}$ is the sum of the expected background contributions, $\mathcal{F}_{\rm res}$ is a resolution factor that accounts for migrations in to and out of the fiducial region, $\varepsilon$ is the efficiency to reconstruct and select the signal events, and $\mathcal{L}$ is the integrated luminosity of the data sample.

The instantaneous luminosity is measured continuously during the acquisition of physics data by recording the rates of several selected reference processes. The cross-section of these processes is measured during a dedicated data-taking calibration period, using a beam-gas imaging method specific to the LHCb detector~\cite{FerroLuzzi:2005em,LHCb-PAPER-2014-047}, which gives a luminosity calibration with an uncertainty of 3.9\%. The integrated luminosity of the data sample used, $\mathcal{L}$, is obtained from the accumulated counts of the calibrated rates and is determined to be $1.93\pm0.07\invfb$.

The event reconstruction and selection efficiency, $\varepsilon$, can be further divided into eight components
\begin{equation}
\varepsilon = \varepsilon^{\rm rec}_{\mu}\cdot\varepsilon^{\rm id}_{\mu}\cdot\varepsilon^{\rm trg}\cdot\varepsilon^{\rm rec}_{e}\cdot\varepsilon^{\rm id}_{e}\cdot\varepsilon^{\rm jet}\cdot\varepsilon^{\rm tag}\cdot\varepsilon^{\rm sel}
\end{equation}
where the equation is ordered from left to right such that, for each component, the efficiency is evaluated for candidates passing the stages to the left. The efficiencies to reconstruct and identify the muon candidate are given by $\varepsilon^{\rm rec}_{\mu}$ and $\varepsilon^{\rm id}_{\mu}$ respectively, while $\varepsilon^{\rm trg}$ refers to the efficiency to trigger the event on the muon candidate. The efficiencies to reconstruct and identify the electron candidate are given by $\varepsilon^{\rm rec}_{e}$ and $\varepsilon^{\rm id}_{e}$ respectively. The efficiency to reconstruct and tag the jet are given by $\varepsilon^{\rm jet}$ and $\varepsilon^{\rm tag}$ respectively, and the efficiency of the additional selection requirements is given by $\varepsilon^{\rm sel}$.

The efficiencies to reconstruct, identify, and trigger the muon candidate are determined from simulation, where the simulated \pythiapp sample is weighted in the muon \pt and $\eta$ to match NLO predictions from \amcatnlo. Additionally, corrections are applied as a function of the muon \pt and $\eta$ to account for observed differences in the efficiency between data and simulation. The corrections are obtained using a tag-and-probe method on $Z\to\mu^+\mu^-$ events, where one of the muons, the tag, is required to have triggered the event and be fully reconstructed and identified, and a probe is selected that represents the other muon and acts as an unbiased estimator of the efficiency, using similar techniques as those used in Ref.~\cite{LHCB-PAPER-2016-021}. In the case of the reconstruction efficiency, the probe is a track reconstructed using the muon stations and information from tracking detectors not used in the primary track reconstruction algorithms. For the identification efficiency, the probe is a fully reconstructed particle with no identification requirements applied, and for the trigger efficiency, it is a fully reconstructed and identified muon. The uncertainty is determined by combining the statistical uncertainty due to the size of the simulated sample, the uncertainties on the correction factors,  and the difference between the efficiencies obtained with and without the NLO weighting.

%The efficiency to reconstruct and identify the muon candidate using a similar method. For the reconstruction efficiency, the tag-and-probe is performed using a fully reconstructed tag muon, and a probe track reconstructed using only information from the TT and the muon stations which provides an unbiased estimate of the tracking efficiency. For the muon identification efficiency, the probe is a fully reconstructed particle with no identification requirements applied, but with more stringent requirements placed on the tag muon and the event to produce a high purity sample. The kinematics are again reweighted to match the NLO predictions. The identification efficiency is determined using a tag-and-probe method on $Z\to ee$ events, where the tag is a fully reconstructed and identified electron, and the probe is a track.

The efficiencies related to the reconstruction and identification of the electron are again obtained from \pythiapp simulation, weighted as a function of the electron \pt and $\eta$ to match NLO predictions from \amcatnlo. Data-driven studies of these efficiencies are made again using similar techniques to Ref.~\cite{LHCB-PAPER-2016-021}. Corrections to the electron identification efficiency are obtained using a tag-and-probe method on recorded $Z\to e^+e^-$ decays, where the probe is a reconstructed particle with no identification requirements applied. For the electron reconstruction efficiency, no corrections are applied to the simulated sample, but a systematic uncertainty related to the potential mismodelling in simulation is determined by comparing the ratio of the number of $Z\to e^{+}e^{-}$ events where either the two electrons are fully reconstructed, or where one of the electrons is only reconstructed as a deposit in the electromagnetic calorimeter. The statistical uncertainty due to the limited size of the simulated sample is combined with the uncertainty on the corrections and the difference in the efficiency before and after weighting to determine the total uncertainty.

The jet reconstruction and tagging efficiencies are determined directly from \pythiapp simulation with NLO weighting in the jet \pt and $\eta$. A systematic uncertainty on the reconstruction efficiency is determined by comparing the variation in the yield of $Z\to\mu^+\mu^-$ events containing a jet in data and simulation when the quality requirements on the jet are varied, and taking the difference as the uncertainty. The jet tagging efficiency was previously determined to be modelled accurately in simulation to a level of 10\% using data collected in Run 1~\cite{LHCb-PAPER-2015-016}. The level of agreement is evaluated in the data sample used in this paper by comparing the tagging efficiency of jets containing fully reconstructed $B^{\pm}\to\jpsi K^{\pm}$ decays in data and simulation, where the \jpsi meson is reconstructed through its decay to a muon pair. The signal yield is determined by a fit to the invariant mass of the reconstructed $B$ hadron before and after the tagging requirement is applied. A similar level of agreement between data and simulation is observed as in the previous studies, and consequently the same uncertainty is applied.

The selection efficiency refers to the efficiency of the isolation and impact parameter requirements applied to the leptons. The impact parameter efficiency is obtained from \pythiapp simulation where the impact parameter distribution is smeared using factors obtained from a comparison of $Z\to\mu^+\mu^-$ events in data and simulation. The difference between the efficiency obtained before and after the smearing procedure is taken as a systematic uncertainty. The efficiency of the isolation requirement is obtained directly from \pythiapp simulation, with a systematic uncertainty applied to account for differences in data and simulation. The difference in the efficiency of the requirement when applied to $Z\to\mu^+\mu^-$ events in data and simulation is taken as a systematic uncertainty to account for differences due to the underlying event and contributions from additional $pp$ interactions. An additional contribution to the systematic uncertainty, due to possible mismodelling of the $t\bar{t}$ process, is determined as the maximum difference in the efficiency in simulated $t\bar{t}$ events with different jet multiplicities. A summary of the efficiencies and their uncertainties is given in Table~\ref{tab:efficiencies}.

The largest contribution to $\mathcal{F}_{\rm res}$ arises from the momentum resolution of the electron due to bremsstrahlung. A scaling factor is applied to the electron momentum obtained in simulation to better match the \pt spectrum of electrons in $Z\to e^+e^-$ events in data. As the jet resolution also contributes, the level of agreement between data and simulation is evaluated by selecting events containing a $Z$ boson and a jet which are azimuthally opposed. As the ratio of the \pt of the jet and the \Z boson in these events is expected to be close to unity, the width of the distribution gives an estimate of the jet resolution. Corrections are obtained that are used to smear the reconstructed jet energy in simulation. The resolution factor is determined from simulation with both of these corrections applied, and the difference between the computed values before and after the corrections taken as a systematic uncertainty. The resolution factor is determined to be 1.207 $\pm$ 0.006.
\begin{table}
\caption{The efficiency to fully reconstruct and identify the candidates.}
\label{tab:efficiencies}
\begin{center}
\begin{tabular}{l c}
Source & Efficiency \\
\hline
trigger                                & 0.811 $\pm$ 0.016 \\
muon reconstruction          & 0.930 $\pm$ 0.010 \\
electron reconstruction       & 0.916 $\pm$ 0.026\\
muon identification  & 0.978 $\pm$ 0.008\\
electron identification      & 0.918 $\pm$ 0.012 \\
jet reconstruction  & 0.975 $\pm$ 0.016\\
event selection         & 0.564 $\pm$ 0.023  \\
jet tagging        & 0.556 $\pm$ 0.056\\
\hline
total  & 0.190 $\pm$ 0.022
\end{tabular}
\end{center}
\end{table}

% !TEX root = main.tex                                                                                                                                                   
% $Id: introduction.tex 87303 2016-02-08 13:44:29Z lafferty $

\section{Results}
\label{sec:results}
\begin{figure}
\begin{center}
{\setlength{\fboxrule}{1pt}%
\fbox{\includegraphics[width=0.8\textwidth]{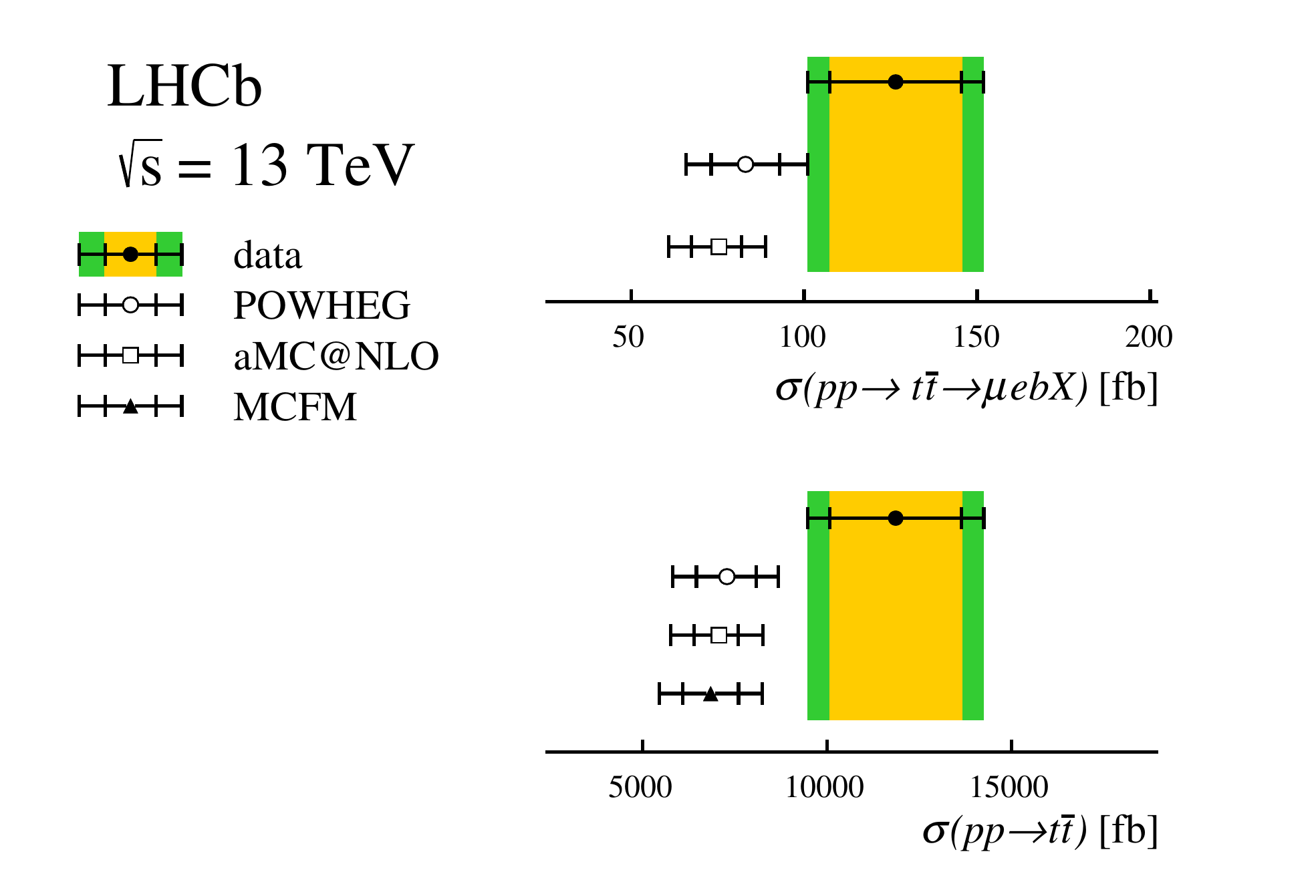}}
}
\end{center}
\caption{Graphical comparison of the measured cross-sections with the predictions from the \amcatnlo, \powheg and \mcfm generators. For the data, the inner error band represents the statistical uncertainty, and the outer the total, while for the theoretical predictions, the inner band represents the scale uncertainty and the outer represents the total. The prediction is shown (above) for the muon, electron and jet fiducial, and (below) for the top quark fiducial region. }
\label{fig:xsec_result}
\end{figure}

Using the formula and inputs described in the previous sections, the cross-section in the fiducial region defined by the \pt, pseudorapidity, and $\Delta R$ requirements placed on the leptons and the \bjet is determined to be
\begin{linenomath}
\begin{equation*}
\sigma_{t\bar{t}}= 126\pm19\,(\mathrm{stat})\pm16\,(\mathrm{syst})\pm5\,(\mathrm{lumi})\,\,\mathrm{ fb}\end{equation*}
\end{linenomath}
where the first uncertainty is statistical, the second is systematic, and the third is due to the luminosity determination. This cross-section is compared to the theoretical predictions obtained from \amcatnlo, \powheg and \mcfm. Three different sources of uncertainty are considered on the theoretical predictions: uncertainties due to the description of the PDFs ($\delta_{\rm PDF}$), the uncertainty due to the choice of renormalisation and factorisation scales ($\delta_{\rm scale}$), and the uncertainty on the value of the strong coupling constant used in the calculation ($\delta_{\as}$). The uncertainty due to the choice of the top quark mass is expected to be small and is not considered further. The total theoretical uncertainty, $\delta_{\rm theory}$, is determined by combining the individual uncertainties according to the formula $\delta_{\rm theory} = \sqrt{\delta_{\rm PDF}^{2} + \delta_{\as}^2} + \delta_{\rm scale}$~\cite{Dittmaier:2011ti}.
A comparison of the measured cross-section with the predictions is shown in Fig.~\ref{fig:xsec_result}. The result is shown in the case where the fiducial requirements are placed on the final state muon, electron and \bjet, and where the fiducial requirements are placed on the top quarks, where the top quarks are defined at parton level after QCD radiation. The latter fiducial region requires that both top quarks have a rapidity between 2.0 and 5.0. The measured cross-section in the top quark fiducial region is obtained by extrapolating from the measured fiducial region using predictions from \amcatnlo, which contributes an additional 1.5\% uncertainty to the measurement uncertainty, evaluated using the same techniques as for the theoretical predictions described earlier. The cross-section in the top quark fiducial region is additionally compared to predictions from \mcfm. The measured cross-section in both fiducial regions is seen to be in agreement with the predictions.
\begin{table}
\caption{Summary of the systematic uncertainties on the measurement of the $t\bar{t}$ cross-section in the fiducial region defined by the requirements placed on the leptons and the \bjet, expressed as a percentage of the measured cross-section. An additional uncertainty due to \ttbar modelling is considered when extrapolating to the top quark fiducial region.}
\label{tab:systs}
\begin{center}
\begin{tabular}{l c}
Source & \% \\
\hline
trigger & \phantom{1}2.0 \\
muon reconstruction & \phantom{1}1.1 \\
electron reconstruction & \phantom{1}2.8 \\
muon identification & \phantom{1}0.8  \\
electron identification & \phantom{1}1.3 \\
jet reconstruction & \phantom{1}1.6\\
event selection & \phantom{1}4.0 \\
jet tagging & 10.0 \\
background & \phantom{1}5.1 \\
resolution factor & \phantom{1}0.5 \\
\hline
total & 12.7
\end{tabular}
\end{center}
\end{table}
A summary of the systematic uncertainties contributing to the measurement is given in Table~\ref{tab:systs}. The dominant systematic uncertainty is due to the knowledge of the jet tagging efficiency.

 \section{Conclusion}
 
\label{sec:conclusions}
 The cross-section for top quark pair production in the forward region at LHCb has been measured using the $\mu e b$ final state, where the presence of a  muon, electron and \bjet are used to identify \ttbar candidates. The cross-section is measured in two fiducial regions: where fiducial requirements are placed on the final state objects, and where fiducial requirements are placed on the top quarks themselves. The latter fiducial cross-section is obtained by scaling the former by an extrapolation factor obtained from \amcatnlo. The measurement precision of 20\% is comparable to prior measurements of top quark production at LHCb in Run 1 and to the precision of the theoretical predictions. The measurement uncertainty is dominated by the statistical precision of the data sample and the knowledge of the $b$-tagging efficiency.
 
The final state presented here is selected with a high purity with respect to measurements in other final states. While a number of systematic uncertainties contributing to the measurement are uncorrelated with measurements in other final states, future reductions of the uncertainty on the $b$-tagging efficiency, common to all final states, will be required in order to fully exploit their complementarity. With the increase in the data-taking capabilities of the LHCb detector in future upgrades~\cite{LHCb-TDR-012,LHCb-PII-EoI}, measurements in the $\mu e b$ final state will no longer be statistically limited, and have the potential to achieve the highest precision on the measurement of the \ttbar production cross-section at LHCb.

% Comment this in for paper darfts; do not include this in analysis note and conference reports
\section*{Acknowledgements}
%
% These Acknowledgements valid from 6-Dec-2017
%
\noindent We express our gratitude to our colleagues in the CERN
accelerator departments for the excellent performance of the LHC. We
thank the technical and administrative staff at the LHCb
institutes. We acknowledge support from CERN and from the national
agencies: CAPES, CNPq, FAPERJ and FINEP (Brazil); MOST and NSFC
(China); CNRS/IN2P3 (France); BMBF, DFG and MPG (Germany); INFN
(Italy); NWO (The Netherlands); MNiSW and NCN (Poland); MEN/IFA
(Romania); MinES and FASO (Russia); MinECo (Spain); SNSF and SER
(Switzerland); NASU (Ukraine); STFC (United Kingdom); NSF (USA).  We
acknowledge the computing resources that are provided by CERN, IN2P3
(France), KIT and DESY (Germany), INFN (Italy), SURF (The
Netherlands), PIC (Spain), GridPP (United Kingdom), RRCKI and Yandex
LLC (Russia), CSCS (Switzerland), IFIN-HH (Romania), CBPF (Brazil),
PL-GRID (Poland) and OSC (USA). We are indebted to the communities
behind the multiple open-source software packages on which we depend.
Individual groups or members have received support from AvH Foundation
(Germany), EPLANET, Marie Sk\l{}odowska-Curie Actions and ERC
(European Union), ANR, Labex P2IO and OCEVU, and R\'{e}gion
Auvergne-Rh\^{o}ne-Alpes (France), RFBR, RSF and Yandex LLC (Russia),
GVA, XuntaGal and GENCAT (Spain), Herchel Smith Fund, the Royal
Society, the English-Speaking Union and the Leverhulme Trust (United
Kingdom).

\addcontentsline{toc}{section}{References}
\setboolean{inbibliography}{true}
\bibliographystyle{LHCb}
\bibliography{main,LHCb-PAPER,LHCb-CONF,LHCb-DP,LHCb-TDR}

\ifx\mcitethebibliography\mciteundefinedmacro
\PackageError{LHCb.bst}{mciteplus.sty has not been loaded}
{This bibstyle requires the use of the mciteplus package.}\fi
\providecommand{\href}[2]{#2}
\begin{mcitethebibliography}{10}
\mciteSetBstSublistMode{n}
\mciteSetBstMaxWidthForm{subitem}{\alph{mcitesubitemcount})}
\mciteSetBstSublistLabelBeginEnd{\mcitemaxwidthsubitemform\space}
{\relax}{\relax}

\bibitem{Gauld:2013aja}
R.~Gauld, \ifthenelse{\boolean{articletitles}}{\emph{{Feasibility of top quark
  measurements at LHCb and constraints on the large-$x$ gluon PDF}},
  }{}\href{http://dx.doi.org/10.1007/JHEP02(2014)126}{JHEP \textbf{02} (2014)
  126}, \href{http://arxiv.org/abs/1311.1810}{{\normalfont\ttfamily
  arXiv:1311.1810}}\relax
\mciteBstWouldAddEndPuncttrue
\mciteSetBstMidEndSepPunct{\mcitedefaultmidpunct}
{\mcitedefaultendpunct}{\mcitedefaultseppunct}\relax
\EndOfBibitem
\bibitem{Kagan:2011yx}
A.~L. Kagan, J.~F. Kamenik, G.~Perez, and S.~Stone,
  \ifthenelse{\boolean{articletitles}}{\emph{{Top LHCb physics}},
  }{}\href{http://dx.doi.org/10.1103/PhysRevLett.107.082003}{Phys.\ Rev.\
  Lett.\  \textbf{107} (2011) 082003},
  \href{http://arxiv.org/abs/1103.3747}{{\normalfont\ttfamily
  arXiv:1103.3747}}\relax
\mciteBstWouldAddEndPuncttrue
\mciteSetBstMidEndSepPunct{\mcitedefaultmidpunct}
{\mcitedefaultendpunct}{\mcitedefaultseppunct}\relax
\EndOfBibitem
\bibitem{Gauld:2014pxa}
R.~Gauld, \ifthenelse{\boolean{articletitles}}{\emph{{Leptonic top-quark
  asymmetry predictions at LHCb}},
  }{}\href{http://dx.doi.org/10.1103/PhysRevD.91.054029}{Phys.\ Rev.\
  \textbf{D91} (2015) 054029},
  \href{http://arxiv.org/abs/1409.8631}{{\normalfont\ttfamily
  arXiv:1409.8631}}\relax
\mciteBstWouldAddEndPuncttrue
\mciteSetBstMidEndSepPunct{\mcitedefaultmidpunct}
{\mcitedefaultendpunct}{\mcitedefaultseppunct}\relax
\EndOfBibitem
\bibitem{LHCb-PAPER-2015-022}
LHCb collaboration, R.~Aaij {\em et~al.},
  \ifthenelse{\boolean{articletitles}}{\emph{{First observation of top quark
  production in the forward region}},
  }{}\href{http://dx.doi.org/10.1103/PhysRevLett.115.112001}{Phys.\ Rev.\
  Lett.\  \textbf{115} (2015) 112001},
  \href{http://arxiv.org/abs/1506.00903}{{\normalfont\ttfamily
  arXiv:1506.00903}}\relax
\mciteBstWouldAddEndPuncttrue
\mciteSetBstMidEndSepPunct{\mcitedefaultmidpunct}
{\mcitedefaultendpunct}{\mcitedefaultseppunct}\relax
\EndOfBibitem
\bibitem{LHCb-PAPER-2016-038}
LHCb collaboration, R.~Aaij {\em et~al.},
  \ifthenelse{\boolean{articletitles}}{\emph{{Measurement of the $\ttbar$, $W +
  \bbbar$ and $W + \ccbar$ production cross sections in $\proton\proton$
  collisions at $\sqrt{s}=8$\,TeV}},
  }{}\href{http://dx.doi.org/10.1016/j.physletb.2017.01.044}{Phys.\ Lett.\
  \textbf{B767} (2017) 110},
  \href{http://arxiv.org/abs/1610.08142}{{\normalfont\ttfamily
  arXiv:1610.08142}}\relax
\mciteBstWouldAddEndPuncttrue
\mciteSetBstMidEndSepPunct{\mcitedefaultmidpunct}
{\mcitedefaultendpunct}{\mcitedefaultseppunct}\relax
\EndOfBibitem
\bibitem{Campbell:2000bg}
J.~M. Campbell and R.~K. Ellis,
  \ifthenelse{\boolean{articletitles}}{\emph{{Radiative corrections to Z b
  anti-b production}},
  }{}\href{http://dx.doi.org/10.1103/PhysRevD.62.114012}{Phys.\ Rev.\
  \textbf{D62} (2000) 114012},
  \href{http://arxiv.org/abs/hep-ph/0006304}{{\normalfont\ttfamily
  arXiv:hep-ph/0006304}}\relax
\mciteBstWouldAddEndPuncttrue
\mciteSetBstMidEndSepPunct{\mcitedefaultmidpunct}
{\mcitedefaultendpunct}{\mcitedefaultseppunct}\relax
\EndOfBibitem
\bibitem{Lai:2010vv}
H.-L. Lai {\em et~al.}, \ifthenelse{\boolean{articletitles}}{\emph{{New parton
  distributions for collider physics}},
  }{}\href{http://dx.doi.org/10.1103/PhysRevD.82.074024}{Phys.\ Rev.\
  \textbf{D82} (2010) 074024},
  \href{http://arxiv.org/abs/1007.2241}{{\normalfont\ttfamily
  arXiv:1007.2241}}\relax
\mciteBstWouldAddEndPuncttrue
\mciteSetBstMidEndSepPunct{\mcitedefaultmidpunct}
{\mcitedefaultendpunct}{\mcitedefaultseppunct}\relax
\EndOfBibitem
\bibitem{Gauld:1557385}
R.~Gauld, \ifthenelse{\boolean{articletitles}}{\emph{{Measuring top quark
  production asymmetries at LHCb}}, }{}
  \href{http://cdsweb.cern.ch/search?p=LHCb-PUB-2013-009&f=reportnumber&action_search=Search&c=LHCb+Notes}
  {LHCb-PUB-2013-009}\relax
\mciteBstWouldAddEndPuncttrue
\mciteSetBstMidEndSepPunct{\mcitedefaultmidpunct}
{\mcitedefaultendpunct}{\mcitedefaultseppunct}\relax
\EndOfBibitem
\bibitem{Alves:2008zz}
LHCb collaboration, A.~A. Alves~Jr.\ {\em et~al.},
  \ifthenelse{\boolean{articletitles}}{\emph{{The \lhcb detector at the LHC}},
  }{}\href{http://dx.doi.org/10.1088/1748-0221/3/08/S08005}{JINST \textbf{3}
  (2008) S08005}\relax
\mciteBstWouldAddEndPuncttrue
\mciteSetBstMidEndSepPunct{\mcitedefaultmidpunct}
{\mcitedefaultendpunct}{\mcitedefaultseppunct}\relax
\EndOfBibitem
\bibitem{LHCb-DP-2014-002}
LHCb collaboration, R.~Aaij {\em et~al.},
  \ifthenelse{\boolean{articletitles}}{\emph{{LHCb detector performance}},
  }{}\href{http://dx.doi.org/10.1142/S0217751X15300227}{Int.\ J.\ Mod.\ Phys.\
  \textbf{A30} (2015) 1530022},
  \href{http://arxiv.org/abs/1412.6352}{{\normalfont\ttfamily
  arXiv:1412.6352}}\relax
\mciteBstWouldAddEndPuncttrue
\mciteSetBstMidEndSepPunct{\mcitedefaultmidpunct}
{\mcitedefaultendpunct}{\mcitedefaultseppunct}\relax
\EndOfBibitem
\bibitem{Sjostrand:2007gs}
T.~Sj\"{o}strand, S.~Mrenna, and P.~Skands,
  \ifthenelse{\boolean{articletitles}}{\emph{{A brief introduction to PYTHIA
  8.1}}, }{}\href{http://dx.doi.org/10.1016/j.cpc.2008.01.036}{Comput.\ Phys.\
  Commun.\  \textbf{178} (2008) 852},
  \href{http://arxiv.org/abs/0710.3820}{{\normalfont\ttfamily
  arXiv:0710.3820}}\relax
\mciteBstWouldAddEndPuncttrue
\mciteSetBstMidEndSepPunct{\mcitedefaultmidpunct}
{\mcitedefaultendpunct}{\mcitedefaultseppunct}\relax
\EndOfBibitem
\bibitem{Sjostrand:2006za}
T.~Sj\"{o}strand, S.~Mrenna, and P.~Skands,
  \ifthenelse{\boolean{articletitles}}{\emph{{PYTHIA 6.4 physics and manual}},
  }{}\href{http://dx.doi.org/10.1088/1126-6708/2006/05/026}{JHEP \textbf{05}
  (2006) 026}, \href{http://arxiv.org/abs/hep-ph/0603175}{{\normalfont\ttfamily
  arXiv:hep-ph/0603175}}\relax
\mciteBstWouldAddEndPuncttrue
\mciteSetBstMidEndSepPunct{\mcitedefaultmidpunct}
{\mcitedefaultendpunct}{\mcitedefaultseppunct}\relax
\EndOfBibitem
\bibitem{Allison:2006ve}
Geant4 collaboration, J.~Allison {\em et~al.},
  \ifthenelse{\boolean{articletitles}}{\emph{{Geant4 developments and
  applications}}, }{}\href{http://dx.doi.org/10.1109/TNS.2006.869826}{IEEE
  Trans.\ Nucl.\ Sci.\  \textbf{53} (2006) 270}\relax
\mciteBstWouldAddEndPuncttrue
\mciteSetBstMidEndSepPunct{\mcitedefaultmidpunct}
{\mcitedefaultendpunct}{\mcitedefaultseppunct}\relax
\EndOfBibitem
\bibitem{LHCb-PROC-2011-006}
M.~Clemencic {\em et~al.}, \ifthenelse{\boolean{articletitles}}{\emph{{The
  \lhcb simulation application, Gauss: Design, evolution and experience}},
  }{}\href{http://dx.doi.org/10.1088/1742-6596/331/3/032023}{{J.\ Phys.\ Conf.\
  Ser.\ } \textbf{331} (2011) 032023}\relax
\mciteBstWouldAddEndPuncttrue
\mciteSetBstMidEndSepPunct{\mcitedefaultmidpunct}
{\mcitedefaultendpunct}{\mcitedefaultseppunct}\relax
\EndOfBibitem
\bibitem{Nason:2004rx}
P.~Nason, \ifthenelse{\boolean{articletitles}}{\emph{{A new method for
  combining NLO QCD with shower Monte Carlo algorithms}},
  }{}\href{http://dx.doi.org/10.1088/1126-6708/2004/11/040}{JHEP \textbf{11}
  (2004) 040}, \href{http://arxiv.org/abs/hep-ph/0409146}{{\normalfont\ttfamily
  arXiv:hep-ph/0409146}}\relax
\mciteBstWouldAddEndPuncttrue
\mciteSetBstMidEndSepPunct{\mcitedefaultmidpunct}
{\mcitedefaultendpunct}{\mcitedefaultseppunct}\relax
\EndOfBibitem
\bibitem{Frixione:2007vw}
S.~Frixione, P.~Nason, and C.~Oleari,
  \ifthenelse{\boolean{articletitles}}{\emph{{Matching NLO QCD computations
  with parton shower simulations: the POWHEG method}},
  }{}\href{http://dx.doi.org/10.1088/1126-6708/2007/11/070}{JHEP \textbf{11}
  (2007) 070}, \href{http://arxiv.org/abs/0709.2092}{{\normalfont\ttfamily
  arXiv:0709.2092}}\relax
\mciteBstWouldAddEndPuncttrue
\mciteSetBstMidEndSepPunct{\mcitedefaultmidpunct}
{\mcitedefaultendpunct}{\mcitedefaultseppunct}\relax
\EndOfBibitem
\bibitem{Alioli:2010xd}
S.~Alioli, P.~Nason, C.~Oleari, and E.~Re,
  \ifthenelse{\boolean{articletitles}}{\emph{{A general framework for
  implementing NLO calculations in shower Monte Carlo programs: The POWHEG
  BOX}}, }{}\href{http://dx.doi.org/10.1007/JHEP06(2010)043}{JHEP \textbf{06}
  (2010) 043}, \href{http://arxiv.org/abs/1002.2581}{{\normalfont\ttfamily
  arXiv:1002.2581}}\relax
\mciteBstWouldAddEndPuncttrue
\mciteSetBstMidEndSepPunct{\mcitedefaultmidpunct}
{\mcitedefaultendpunct}{\mcitedefaultseppunct}\relax
\EndOfBibitem
\bibitem{Alwall:2014hca}
J.~Alwall {\em et~al.}, \ifthenelse{\boolean{articletitles}}{\emph{{The
  automated computation of tree-level and next-to-leading order differential
  cross sections, and their matching to parton shower simulations}},
  }{}\href{http://dx.doi.org/10.1007/JHEP07(2014)079}{JHEP \textbf{07} (2014)
  079}, \href{http://arxiv.org/abs/1405.0301}{{\normalfont\ttfamily
  arXiv:1405.0301}}\relax
\mciteBstWouldAddEndPuncttrue
\mciteSetBstMidEndSepPunct{\mcitedefaultmidpunct}
{\mcitedefaultendpunct}{\mcitedefaultseppunct}\relax
\EndOfBibitem
\bibitem{Campbell:1999ah}
J.~M. Campbell and R.~K. Ellis, \ifthenelse{\boolean{articletitles}}{\emph{{An
  update on vector boson pair production at hadron colliders}},
  }{}\href{http://dx.doi.org/10.1103/PhysRevD.60.113006}{Phys.\ Rev.\
  \textbf{D60} (1999) 113006},
  \href{http://arxiv.org/abs/hep-ph/9905386}{{\normalfont\ttfamily
  arXiv:hep-ph/9905386}}\relax
\mciteBstWouldAddEndPuncttrue
\mciteSetBstMidEndSepPunct{\mcitedefaultmidpunct}
{\mcitedefaultendpunct}{\mcitedefaultseppunct}\relax
\EndOfBibitem
\bibitem{Campbell:2011bn}
J.~M. Campbell, R.~K. Ellis, and C.~Williams,
  \ifthenelse{\boolean{articletitles}}{\emph{{Vector boson pair production at
  the LHC}}, }{}\href{http://dx.doi.org/10.1007/JHEP07(2011)018}{JHEP
  \textbf{07} (2011) 018},
  \href{http://arxiv.org/abs/1105.0020}{{\normalfont\ttfamily
  arXiv:1105.0020}}\relax
\mciteBstWouldAddEndPuncttrue
\mciteSetBstMidEndSepPunct{\mcitedefaultmidpunct}
{\mcitedefaultendpunct}{\mcitedefaultseppunct}\relax
\EndOfBibitem
\bibitem{Frixione:2007zp}
S.~Frixione, E.~Laenen, P.~Motylinski, and B.~R. Webber,
  \ifthenelse{\boolean{articletitles}}{\emph{{Angular correlations of lepton
  pairs from vector boson and top quark decays in Monte Carlo simulations}},
  }{}\href{http://dx.doi.org/10.1088/1126-6708/2007/04/081}{JHEP \textbf{04}
  (2007) 081}, \href{http://arxiv.org/abs/hep-ph/0702198}{{\normalfont\ttfamily
  arXiv:hep-ph/0702198}}\relax
\mciteBstWouldAddEndPuncttrue
\mciteSetBstMidEndSepPunct{\mcitedefaultmidpunct}
{\mcitedefaultendpunct}{\mcitedefaultseppunct}\relax
\EndOfBibitem
\bibitem{Artoisenet:2012st}
P.~Artoisenet, R.~Frederix, O.~Mattelaer, and R.~Rietkerk,
  \ifthenelse{\boolean{articletitles}}{\emph{{Automatic spin-entangled decays
  of heavy resonances in Monte Carlo simulations}},
  }{}\href{http://dx.doi.org/10.1007/JHEP03(2013)015}{JHEP \textbf{03} (2013)
  015}, \href{http://arxiv.org/abs/1212.3460}{{\normalfont\ttfamily
  arXiv:1212.3460}}\relax
\mciteBstWouldAddEndPuncttrue
\mciteSetBstMidEndSepPunct{\mcitedefaultmidpunct}
{\mcitedefaultendpunct}{\mcitedefaultseppunct}\relax
\EndOfBibitem
\bibitem{Re:2010bp}
E.~Re, \ifthenelse{\boolean{articletitles}}{\emph{{Single-top Wt-channel
  production matched with parton showers using the POWHEG method}},
  }{}\href{http://dx.doi.org/10.1140/epjc/s10052-011-1547-z}{Eur.\ Phys.\ J.\
  \textbf{C71} (2011) 1547},
  \href{http://arxiv.org/abs/1009.2450}{{\normalfont\ttfamily
  arXiv:1009.2450}}\relax
\mciteBstWouldAddEndPuncttrue
\mciteSetBstMidEndSepPunct{\mcitedefaultmidpunct}
{\mcitedefaultendpunct}{\mcitedefaultseppunct}\relax
\EndOfBibitem
\bibitem{Ball:2014uwa}
NNPDF collaboration, R.~D. Ball {\em et~al.},
  \ifthenelse{\boolean{articletitles}}{\emph{{Parton distributions for the LHC
  Run II}}, }{}\href{http://dx.doi.org/10.1007/JHEP04(2015)040}{JHEP
  \textbf{04} (2015) 040},
  \href{http://arxiv.org/abs/1410.8849}{{\normalfont\ttfamily
  arXiv:1410.8849}}\relax
\mciteBstWouldAddEndPuncttrue
\mciteSetBstMidEndSepPunct{\mcitedefaultmidpunct}
{\mcitedefaultendpunct}{\mcitedefaultseppunct}\relax
\EndOfBibitem
\bibitem{Cacciari:2011ma}
M.~Cacciari, G.~P. Salam, and G.~Soyez,
  \ifthenelse{\boolean{articletitles}}{\emph{{FastJet user manual}},
  }{}\href{http://dx.doi.org/10.1140/epjc/s10052-012-1896-2}{Eur.\ Phys.\ J.\
  \textbf{C72} (2012) 1896},
  \href{http://arxiv.org/abs/1111.6097}{{\normalfont\ttfamily
  arXiv:1111.6097}}\relax
\mciteBstWouldAddEndPuncttrue
\mciteSetBstMidEndSepPunct{\mcitedefaultmidpunct}
{\mcitedefaultendpunct}{\mcitedefaultseppunct}\relax
\EndOfBibitem
\bibitem{LHCb-PAPER-2015-016}
LHCb collaboration, R.~Aaij {\em et~al.},
  \ifthenelse{\boolean{articletitles}}{\emph{{Identification of beauty and
  charm quark jets at LHCb}},
  }{}\href{http://dx.doi.org/10.1088/1748-0221/10/06/P06013}{JINST \textbf{10}
  (2015) P06013}, \href{http://arxiv.org/abs/1504.07670}{{\normalfont\ttfamily
  arXiv:1504.07670}}\relax
\mciteBstWouldAddEndPuncttrue
\mciteSetBstMidEndSepPunct{\mcitedefaultmidpunct}
{\mcitedefaultendpunct}{\mcitedefaultseppunct}\relax
\EndOfBibitem
\bibitem{FerroLuzzi:2005em}
M.~Ferro-Luzzi, \ifthenelse{\boolean{articletitles}}{\emph{{Proposal for an
  absolute luminosity determination in colliding beam experiments using vertex
  detection of beam-gas interactions}},
  }{}\href{http://dx.doi.org/10.1016/j.nima.2005.07.010}{Nucl.\ Instrum.\
  Meth.\  \textbf{A553} (2005) 388}\relax
\mciteBstWouldAddEndPuncttrue
\mciteSetBstMidEndSepPunct{\mcitedefaultmidpunct}
{\mcitedefaultendpunct}{\mcitedefaultseppunct}\relax
\EndOfBibitem
\bibitem{LHCb-PAPER-2014-047}
LHCb collaboration, R.~Aaij {\em et~al.},
  \ifthenelse{\boolean{articletitles}}{\emph{{Precision luminosity measurements
  at LHCb}}, }{}\href{http://dx.doi.org/10.1088/1748-0221/9/12/P12005}{JINST
  \textbf{9} (2014) P12005},
  \href{http://arxiv.org/abs/1410.0149}{{\normalfont\ttfamily
  arXiv:1410.0149}}\relax
\mciteBstWouldAddEndPuncttrue
\mciteSetBstMidEndSepPunct{\mcitedefaultmidpunct}
{\mcitedefaultendpunct}{\mcitedefaultseppunct}\relax
\EndOfBibitem
\bibitem{LHCB-PAPER-2016-021}
LHCb collaboration, R.~Aaij {\em et~al.},
  \ifthenelse{\boolean{articletitles}}{\emph{{Measurement of the forward $\Z$
  boson production cross-section in $\proton\proton$ collisions at
  $\sqrt{s}=13$\,TeV}},
  }{}\href{http://dx.doi.org/10.1007/JHEP09(2016)136}{JHEP \textbf{09} (2016)
  136}, \href{http://arxiv.org/abs/1607.06495}{{\normalfont\ttfamily
  arXiv:1607.06495}}\relax
\mciteBstWouldAddEndPuncttrue
\mciteSetBstMidEndSepPunct{\mcitedefaultmidpunct}
{\mcitedefaultendpunct}{\mcitedefaultseppunct}\relax
\EndOfBibitem
\bibitem{Dittmaier:2011ti}
LHC Higgs Cross Section Working Group, S.~Dittmaier {\em et~al.},
  \ifthenelse{\boolean{articletitles}}{\emph{{Handbook of LHC Higgs Cross
  Sections: 1. Inclusive Observables}},
  }{}\href{http://arxiv.org/abs/1101.0593}{{\normalfont\ttfamily
  arXiv:1101.0593}}\relax
\mciteBstWouldAddEndPuncttrue
\mciteSetBstMidEndSepPunct{\mcitedefaultmidpunct}
{\mcitedefaultendpunct}{\mcitedefaultseppunct}\relax
\EndOfBibitem
\bibitem{LHCb-TDR-012}
LHCb collaboration, \ifthenelse{\boolean{articletitles}}{\emph{{Framework TDR
  for the LHCb Upgrade: Technical Design Report}}, }{}
  \href{http://cdsweb.cern.ch/search?p=CERN-LHCC-2012-007&f=reportnumber&action_search=Search&c=LHCb+Reports}
  {CERN-LHCC-2012-007}\relax
\mciteBstWouldAddEndPuncttrue
\mciteSetBstMidEndSepPunct{\mcitedefaultmidpunct}
{\mcitedefaultendpunct}{\mcitedefaultseppunct}\relax
\EndOfBibitem
\bibitem{LHCb-PII-EoI}
LHCb collaboration, \ifthenelse{\boolean{articletitles}}{\emph{{Expression of
  Interest for a Phase-II LHCb Upgrade: Opportunities in flavour physics, and
  beyond, in the HL-LHC era}}, }{}
  \href{http://cdsweb.cern.ch/search?p=CERN-LHCC-2017-003&f=reportnumber&action_search=Search&c=LHCb+Reports}
  {CERN-LHCC-2017-003}\relax
\mciteBstWouldAddEndPuncttrue
\mciteSetBstMidEndSepPunct{\mcitedefaultmidpunct}
{\mcitedefaultendpunct}{\mcitedefaultseppunct}\relax
\EndOfBibitem
\end{mcitethebibliography}

\newpage

% Author List ----------------------------                                                                                                                                                                                                                                                                                                
%  You need to get a new author list!                                                                                                                                                                                                                                                                                                    

%\input{LHCb_HD_authorlist_2014-06-20}
 
\newpage
\centerline{\large\bf LHCb collaboration}
\begin{flushleft}
\small
R.~Aaij$^{40}$,
B.~Adeva$^{39}$,
M.~Adinolfi$^{48}$,
Z.~Ajaltouni$^{5}$,
S.~Akar$^{59}$,
J.~Albrecht$^{10}$,
F.~Alessio$^{40}$,
M.~Alexander$^{53}$,
A.~Alfonso~Albero$^{38}$,
S.~Ali$^{43}$
% ...

{\footnotesize \it
% ...
$ ^{5}$Clermont Universit{\'e}, Universit{\'e} Blaise Pascal, CNRS/IN2P3, LPC, Clermont-Ferrand, France\\
% ...
$ ^{10}$Fakult{\"a}t Physik, Technische Universit{\"a}t Dortmund, Dortmund, Germany\\
% ...
$ ^{38}$ICCUB, Universitat de Barcelona, Barcelona, Spain\\
$ ^{39}$Instituto Galego de F{\'\i}sica de Altas Enerx{\'\i}as (IGFAE), Universidade de Santiago de Compostela, Santiago de Compostela, Spain\\
$ ^{40}$European Organization for Nuclear Research (CERN), Geneva, Switzerland\\
% ...
$ ^{43}$Nikhef National Institute for Subatomic Physics, Amsterdam, The Netherlands\\
% ...
$ ^{48}$H.H. Wills Physics Laboratory, University of Bristol, Bristol, United Kingdom\\
% ...
$ ^{53}$School of Physics and Astronomy, University of Glasgow, Glasgow, United Kingdom\\
% ...
$ ^{59}$University of Cincinnati, Cincinnati, OH, United States\\
% ...
}
\end{flushleft}

\end{document}